\documentclass[aps,twocolumn, secnumarabic,amssymb,amsfonts,amsmath,nobibnotes, prl, superscriptaddress,showkeys, floatfix]{revtex4-1}
\usepackage[english]{babel}
\usepackage[utf8]{inputenc}
\usepackage[colorinlistoftodos, color=green!40, prependcaption]{todonotes}

\usepackage{amsthm}
\usepackage{mathtools}

\usepackage{xcolor}
\usepackage{graphicx}
\usepackage[left=23mm,right=13mm,top=35mm,columnsep=15pt]{geometry} 
\usepackage{adjustbox}
\usepackage{placeins}
\usepackage[T1]{fontenc}
\usepackage{lipsum}
\usepackage{csquotes}
\usepackage{hyperref}
\usepackage{cleveref}
\crefname{equation}{Eq.}{Eqs.}
\Crefname{equation}{Equation}{Equations}
\crefname{figure}{Fig.}{Figs.}
\Crefname{figure}{Figure}{Figures}
\crefname{section}{Sec.}{Secs.}
\Crefname{section}{Section}{Sections}
\crefname{appendix}{Appendix}{Apps.}
\Crefname{appendix}{Appendix}{Apps.}
\crefname{paragraph}{Sec.}{Secs.}
\crefname{table}{Table}{Tables}
\usepackage{amsfonts}
\usepackage{physics}

\newcommand{\hhh}{\textbf{H}}

\newcommand{\aaa}{\textbf{a}}
\newcommand{\JJJ}{\textbf{J}}
\newcommand{\ddd}{\textbf{d}}
\newcommand{\bbb}{\textbf{b}}
\newcommand{\sss}{\boldsymbol{\sigma}}
\newcommand{\alphabold}{\boldsymbol{\alpha}}

\newcommand{\jc}[1]{{\color{brown}{[JC: #1]}}}
\newcommand{\comment}[1]{}
\usepackage{color}
\usepackage{soul}

\begin{document}
% \title{Mølmer-Sørensen Gate for Rabi Driven Superconducting Qubits}
\title{Multipartite Entanglement in Rabi Driven Superconducting Qubits}

\author{Marie Lu}
    \email[Correspondence email address: ]{marielu@berkeley.edu}% Your name
    \affiliation{Department of Physics, University of California, Berkeley, California, 94720, USA}
\author{Jean-Loup Ville}
    % \email[Correspondence email address: ]{jlville@berkeley.edu}% Your name
    \affiliation{Department of Physics, University of California, Berkeley, California, 94720, USA}
\author{Joachim Cohen}
    % \email[Correspondence email address: ]{cohenj38@gmail.com}% Your name
    \affiliation{Institut Quantique and D\'epartement de Physique, Universit\'e de Sherbrooke, Sherbrooke, Qu\'ebec, J1K 2R1, Canada}
\author{Alexandru Petrescu}
    % \email[Correspondence email address: ]{tpetresc@gmail.com}% Your name
    \affiliation{Institut Quantique and D\'epartement de Physique, Universit\'e de Sherbrooke, Sherbrooke, Qu\'ebec, J1K 2R1, Canada}
\author{Sydney Schreppler}
    % \email[Correspondence email address: ]{sschreppler@gmail.com}% Your name
    % \affiliation{Microsoft Corp}
    \affiliation{Department of Physics, University of California, Berkeley, California, 94720, USA}
\author{Larry Chen}
    % \email[Correspondence email address: ]{larrychen@berkeley.edu}% Your name
    \affiliation{Department of Physics, University of California, Berkeley, California, 94720, USA}
\author{Christian J\"{u}nger}
    % \email[Correspondence email address: ]{christian.juenger@berkeley.edu}
    \affiliation{Applied Mathematics and Computational Research Division, Lawrence Berkeley National Laboratory, University of California, Berkeley, California, 94720, USA}
\author{Chiara Pelletti}
    % \email[Correspondence email address: ]{chiara@bleximo.com}% Your name
    \affiliation{Bleximo Corp., 701 Heinz Ave, Berkeley, CA 94710, USA}
\author{Alexei Marchenkov}
    % \email[Correspondence email address: ]{alexei.marchenkov@gmail.com}% Your name
    \affiliation{Bleximo Corp., 701 Heinz Ave, Berkeley, CA 94710, USA}
\author{Archan Banerjee}
    % \email[Correspondence email address: ]{archanbanerjee@lbl.gov}% Your name
    \affiliation{Department of Physics, University of California, Berkeley, California, 94720, USA}
\author{William P. Livingston}
    % \email[Correspondence email address: ]{wlivingston@berkeley.edu}% Your name
    \affiliation{Department of Physics, University of California, Berkeley, California, 94720, USA}
\author{John Mark Kreikebaum}
    % \email[Correspondence email address: ]{jmkreikebaum@yahoo.com}% Your name
    \affiliation{Applied Mathematics and Computational Research Division, Lawrence Berkeley National Laboratory, University of California, Berkeley, California, 94720, USA}
\author{David I. Santiago}
    % \email[Correspondence email address: ]{david.i.santiago@lbl.gov}% Your name
    \affiliation{Applied Mathematics and Computational Research Division, Lawrence Berkeley National Laboratory, University of California, Berkeley, California, 94720, USA}
\author{Alexandre Blais}
    % \email[Correspondence email address: ]{alexandre.blais@usherbrooke.ca}% Your name
    % \affiliation{Lawrence Berkeley National Laboratory, University of California, Berkeley, Berkeley CA 94720}
    \affiliation{Institut Quantique and D\'epartement de Physique, Universit\'e de Sherbrooke, Sherbrooke, Qu\'ebec, J1K 2R1, Canada}
    \affiliation{Canadian Institute for Advanced Research, Toronto, M5G1M1 Ontario, Canada}
\author{Irfan Siddiqi}
    % \email[Correspondence email address: ]{irfan_siddiqi@berkeley.edu}% Your name
    \affiliation{Department of Physics, University of California, Berkeley, California, 94720, USA}
    \affiliation{Applied Mathematics and Computational Research Division, Lawrence Berkeley National Laboratory, University of California, Berkeley, California, 94720, USA}
    
\date{\today} % Leave empty to omit a date

\begin{abstract}

\comment{V1: Exploring highly connected networks of qubits is invaluable for implementing various quantum algorithms and simulations. All-to-all connectivity allows for entangling qubits with reduced gate depth. On trapped ion quantum processors, the shared motional degree of freedom is routinely used to simultaneously entangle over a dozen qubits with high fidelity in the  M{\o}lmer-S{\o}rensen gate. Here, we demonstrate the first implementation of a Mølmer-Sørensen-like interaction in circuit quantum electrodynamics through assisted by a shared photonic mode and Rabi driven qubits, which relaxes restrictions on qubit frequencies during fabrication and supports scalability. We achieve a two-qubit entangled state with maximum state fidelity or 97\% in 310 ns and a three-qubit GHZ state with fidelity 90.5\% in 217 ns, and a four-qubit GHZ state with 66\% fidelity. The four-qubit gate is limited by shared resonator losses and the spread of qubit-resonator couplings, which must be addressed for high fidelity operations.}

Exploring highly connected networks of qubits is invaluable for implementing various quantum algorithms and simulations as it allows for entangling qubits with reduced circuit depth. Here, we demonstrate a multi-qubit STAR (Sideband Tone Assisted Rabi driven) gate. Our scheme is inspired by the ion qubit M{\o}lmer-S{\o}rensen gate and is mediated by a shared photonic mode and Rabi-driven superconducting qubits, which relaxes restrictions on qubit frequencies during fabrication and supports scalability. We achieve a two-qubit gate with maximum state fidelity of 95\% in 310 ns, a three-qubit gate with state fidelity 90.5\% in 217 ns, and a four-qubit gate with state fidelity 66\% in 200 ns. Furthermore, we develop a model of the gate that show the four-qubit gate is limited by shared resonator losses and the spread of qubit-resonator couplings, which must be addressed to reach high-fidelity operations.
\end{abstract}

% \keywords{first keyword, second keyword, third keyword}

\maketitle
 
\section{Introduction}
Quantum computers will enable efficient solutions to problems currently intractable for classical computers \cite{chow_2012,grover_1997, brunner_nonlocality_2014,Alvarez_QC}. Among the many models of quantum computation is the gate-based quantum computer, which uses gates to act upon a register of qubits. An algorithm can be broken down into single-qubit operations and two-qubit entangling gates. However, as the leading quantum processors today are limited to 50-100 qubits and each qubit is sensitive to decoherence
(often referred to as NISQ, or noisy intermediate-scale quantum, devices \cite{Preskill_NISQ}), running algorithms with large circuit depth remains difficult. High-fidelity multi-qubit gates are therefore central for reducing circuit depth \cite{brickman_2005,Maslov_2018}. \comment{such as photon qubits \cite{Wang_2018,zhong_2018,omran_2019}, atoms  \cite{graham2022}, cat qubits \cite{Wang_2018,song_2019}, solid state qubits  \cite{Bradley_2019}, and in particular, ion qubits \cite{wangMS_2020, ballance_2016, gaebler_2016, monz_2011, Friis_2018}. }

\comment{In addition to having high fidelity two-qubit gates \cite{mitchell_2021, kandala_2021, stehlik_2021,negirneac_2021, sung_2021_2q}}
Superconducting circuits are a promising platform for quantum computation, due to advances in scaling up microfabrication technology, individual qubit control and readout, and growing qubit coherence time \cite{kjaergaard_2020,nersisyan_2019}. However, generating high-fidelity multi-qubit entanglement on cQED platforms, especially for three and more qubits, remains difficult as multi-qubit gates are limited by connectivity between qubits, noise in controls, and frequency crowding\cite{KrantzOliver_guide, KjaergaardOliver_stateOfPlay}. In particular, the connectivity of superconducting qubits is a design challenge as it is difficult to place $O(n^2)$ qubits in an area that fits $O(n)$ qubits \cite{Maslov_2018}. Processors with nearest-neighbor connectivity can only use cascaded pairwise interactions to generate multi-qubit gates \cite{mooney_2019, gong_2019, wei_2020}, which does not reduce circuit depth. In contrast, tunable qubits coupled to a common bus resonator have the advantage of all-to-all coupling and can entangle by tuning into to near resonance with each other \cite{song10Q_2017}. However, tunable qubits leave a larger footprint on already crowded processors since each qubit requires an additional flux line and generally have shorter coherence times than fixed-frequency qubits, making it more difficult to scale. 

To address these issues, we present the multiqubit STAR (Sideband Tone Assisted Rabi driven) gate on a superconducting quantum processor with four Rabi driven fixed-frequency qubits that are all-to-all connected through a bus resonator. We trade flux tunable qubits for Rabi dressing to harness the advantages of dynamical decoupling, reduce the footprint per qubit, and simplify fabrication restrictions. We demonstrate two-qubit entanglement with maximum state fidelity 95\% (minimum 87\%, average 91\%), three-qubit entanglement with 90.5\% state fidelity, and four-qubit entanglement with fidelity 66\%. In addition, we further characterize the two-qubit gate using quantum process tomography \cite{chuang_QPT, poyatos_QPT}. Moreover, we develop a model of the entangling gate as a blueprint for future scaling. 

\begin{figure*}[ht]
        \includegraphics[width=17cm]{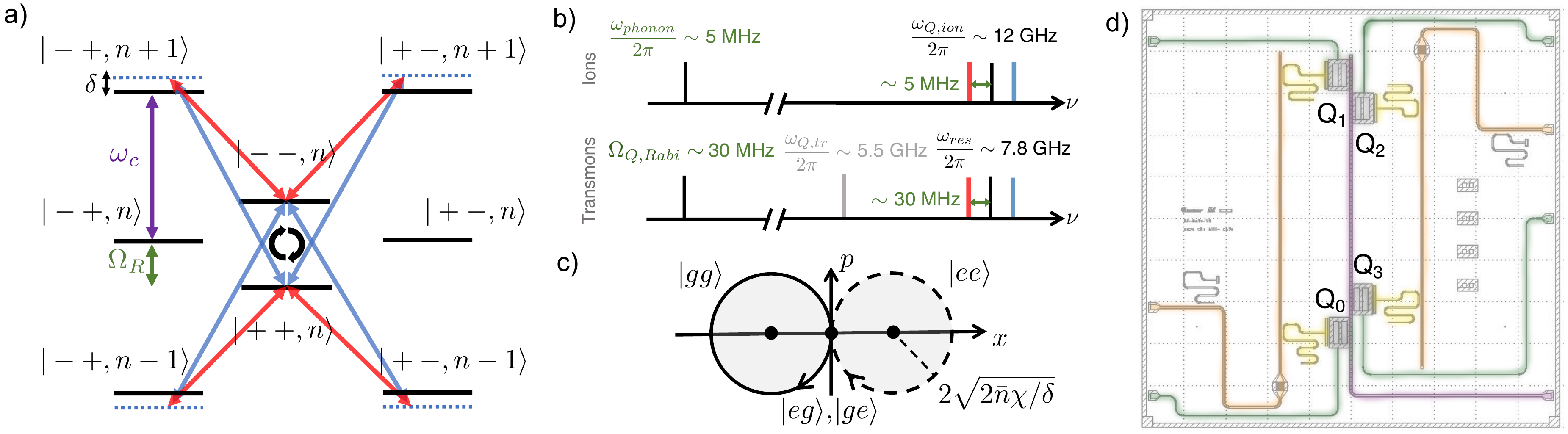}
        \caption{a) \textbf{Experimental implementation on fixed frequency transmons.} Energy level diagram for the Rabi driven qubit states with $n$ photons in the resonator and the sidebands at frequencies $\omega_{r}$ and $\omega_{b}$. Red and blue sidebands drive two photon transitions that generate a population swap between the $\ket{--,n}$ and $\ket{++,n}$ levels. b) Different energy scales used in the gate implementation. Top diagram shows the relevant energies for ion qubits, and the bottom shows the implementation for Rabi driven qubits. c) Trajectories in the $\textbf{x, p}$ plane of the resonator conditioned on the undriven qubit states. The black dots correspond to the displaced vacuums for the corresponding colored states (eigenvalues of $\alphabold$). During the course of the gate, the resonator makes a circle in phase space, entangling and subsequently with the coupled qubits at a rate $1/\delta$. d) Chip design with qubit control lines in green, shared resonator in purple ($\omega_c/2\pi = 7.82$ GHz), readout resonators in yellow ($\omega_{RO}^k/2\pi = $6.4, 6.5, 6.8, 6.6 GHz), and purcell filters in orange. The 4 transmon qubits are the grey rectangles just off of the shared resonator ($\omega_{ge}^k = $ 5.24, 5.37, 5.69, 5.48 GHz).} 
        \label{fig:chip}
\end{figure*}

This gate is reminiscent of the trapped-ion qubit M{\o}lmer-S{\o}rensen gate \cite{MSoriginal1,MSoriginal2,MSoriginal_multi} with a few key differences that can be traced to the distinct energy scales in superconducting circuits and trapped-ion systems. As in the original M{\o}lmer-S{\o}rensen scheme, we use bichromatic sidebands to engineer the interaction between the qubits, as indicated by the red and blue arrows in \cref{fig:chip}a. But while the original ion platform achieves all-to-all coupling through a shared phononic degree of freedom, we exchange this for a coplanar waveguide resonator (CPW). Next, this difference in coupling and qubit platform causes an inversion between the qubit and coupling mode energy scales when comparing the ion and transmon gates
[\cref{fig:chip}b)].
For ions, the high-frequency (GHz) spin degrees of freedom $\sss_i$ are coupled to the low-frequency (MHz) vibrational mode $\aaa$ of the ion chain. The sidebands, which are set at the sum and difference of the frequencies of these two modes, drive the spin degrees of freedom, resulting in resonant hopping terms such as $\sss_i^\pm \aaa$. In the transmon case, the photonic mode is the highest frequency scale, around 8 GHz, similar to the qubit frequency. If we were to generate sidebands at the sum and difference frequencies of the resonator and qubit modes, we would have sidebands that are over 10 GHz apart. These sidebands, that must be sent into the shared CPW resonator, would largely be filtered out by the linewidth of the CPW resonator. Given that our resonators have kHz or MHz intrinsic linewidth, the sidebands must be close in frequency to the shared resonator such that enough power can reach the qubits, requiring a  much lower qubit energy than typical values. 

A key technique used in this gate that both addresses this energy scaling problem and supports its scalability is the addition of a Rabi drive on each qubit. We apply a drive to each participating qubit at its $\ket{g}\longleftrightarrow\ket{e}$ transition frequency to enable a Rabi splitting and form new Rabi dressed qubits with energy levels set by the amplitude of the Rabi drive, $\Omega$, an experimentally tunable parameter \cite{nielsenChuang}. Utilizing these dressed states allows us to relax restrictions on qubit frequencies during fabrication and eliminates the need for tunable qubits. These features support scalability as it reduces the number of lines needed to address each qubit and simplifies the fabrication process. These advantages of Rabi-driven qubits have been further studied and proven to exhibit similar increased coherence properties due to dynamical decoupling in these works \cite{Yan_2013, Yan_2018, Sung_2021_noisespec, von_L_pke_2020, Vool2016, Eddins2018, viola_decoupling}.

\comment{These are just some of the many advantages Rabi-driven qubits present. In the superconducting qubit community, they have also been used as noise sensors \cite{Yan_2013, Yan_2018, Sung_2021_noisespec, von_L_pke_2020}, \comment{for continuous measurement of the $\sss_x$ (transverse component) of a qubit \cite{Vool2016},} and to stroboscopically engineer a longitudinal coupling to enable the use of squeezing for improved measurement \cite{Eddins2018}. Furthermore, the Rabi drive dynamically decouples the qubit from noise and increases the lifetimes relevant to the gate\cite{viola_decoupling}.\jc{I am not sure about the relevance of this paragraph, although I agree that these papers should be cited. Could we say something in the line of "they have been studied and were proven to exhibit similar coherence properties" ?}}
\section{Circuit QED Implementation} \label{sec2}

Experimental implementation details are shown in \cref{fig:chip}. The four-qubit processor in \cref{fig:chip}d displays fixed-frequency transmons that have qubit frequencies $\omega_{ge}^{0,1,2,3}/2\pi = 5.24, 5.37, 5.69, 5.48$ GHz, anharmonicity $\alpha$/2$\pi = -240$ MHz, and dispersive shifts to the common bus mode $\chi_{0,1,2,3}/2\pi = 380, 410, 718, 815$ KHz. The resonator frequency is $\omega_c/2\pi = 7.82$ GHz and has a linewidth $\kappa$ of 180 KHz. This value is dominated by the internal loss of the resonator, as the designed $\kappa_{ext}$ is only 20 KHz. The qubit transitions relevant for the gate are the Rabi split states $\ket{\pm}=(\ket{e}\pm\ket{g})/\sqrt{2}$ generated by applying drives at each qubit frequency, $\omega_{ge}^k$. These are the states displayed in the two-qubit level diagram in \cref{fig:chip}a, where $n$ labels the photon occupation of the shared resonator.  Red and blue sideband tones are applied to the resonator, with frequencies $\omega_{r} = \omega_c - \Omega_R + \delta$ and $\omega_{b} = \omega_c + \Omega_R + \delta$. Here, $\delta$ is the common detuning of the sideband tones. As indicated by the thick black arrows in \cref{fig:chip}a, these sidebands drive two-photon transitions that result in population swaps between $\ket{++,n}$ and $\ket{--,n}$ if the qubits are prepared in either $\ket{++,n}$ or $\ket{--,n}$. We note that one may also initialize the qubits in $\ket{-+,n}$ and generate population swaps with $\ket{+-,n}$.

In the frame rotating at the qubit frequencies and at $\omega_c + \delta$ for an $n$-qubit gate, the dispersive Hamiltonian describing the system is given by \cite{Blais_2004}
\begin{equation}
    \hhh_I = -\delta \aaa^\dagger \aaa+ \sum_{k=1}^{n} (\Omega_R/2)\sss_{x_k}  -\chi_k \sss_{z_k} \aaa^\dagger \aaa  + \hhh_\text{sb}(t).
\end{equation}
Here, $\aaa$ is the annihilation operator of the resonator, and $\sss_{l_k},~l = x,y,z$, are the Pauli matrices of qubit $k$. The effect of the sideband term $\hhh_\text{sb}(t)$ acting on the resonator mode is to displace the field of the resonator such that  $\aaa = \ddd + \alpha(t)$, with $\alpha(t) = \sqrt{2\bar{n}} \cos(\Omega_{R} t + \varphi_\Delta)$, where $\bar{n}$ is the mean resonator photon number due to the two sideband tones, and $\varphi_\Delta$ is the phase difference of the two tones. After performing the displacement transformation and going into the qubit frame at the Rabi frequency [please see Supplement for details], the Hamiltonian reduces to 
\begin{equation}\label{eq:simplifiedHamiltonian}
    \hhh_R = -\delta \ddd^\dagger \ddd -2\sqrt{2\bar{n}}\chi \JJJ_{zy}^{\varphi_\Delta} (\ddd+\ddd^\dagger)  +\hhh_\text{err},
\end{equation}
where $\JJJ_{\varphi_\Delta} = \cos(\varphi_\Delta)\JJJ_{z}-\sin(\varphi_\Delta) \JJJ_{y}$, $\JJJ_{l} = \sum_k \sss_{l_k}/2,~l=x,y,z$ are the generalized spin operators, and $\hhh_\text{err} = \textbf{A}_1 e^{i\Omega_R t }+\textbf{A}_1 e^{2i\Omega_R t }+\textbf{h.c}$ are spurious oscillating terms \cite{Eddins2018}. The effect of these terms can be neglected in the limit of large Rabi frequency, $\delta \sim \chi \sqrt{\bar{n}} < \chi {\bar{n}} \ll \Omega_R$. In this parameter regime, the Hamiltonian of  \cref{eq:simplifiedHamiltonian} can be mapped to the Hamiltonian originally proposed by M{\o}lmer and S{\o}rensen in the context of trapped ions.

Here, we give a brief review of the working principle of the STAR gate. During the gate, the qubits entangle with the resonator field, resulting in a non-trivial operation on the qubits $U = e^{i\frac{\pi}{2}\JJJ_{\varphi_\Delta}^2}$. The origin of this non-trivial phase comes from the qubit-state-dependent paths that the resonator describes in phase-space. To see this, one can cast the Hamiltonian of \cref{eq:simplifiedHamiltonian} in the form $\hhh_R = -\delta (\ddd^\dagger-\alphabold^{*})(\ddd-\alphabold)-\alphabold^2$, with $\alphabold = 2\chi\sqrt{2\bar{n}}\JJJ_{\varphi_\Delta}/\delta$. The first term describes a harmonic oscillator of frequency $\delta$ centered on the qubit-state dependent $\alphabold$, while the second term $\alphabold^2 = 8\bar{n}\chi^2\JJJ_{\varphi_\Delta}^2/\delta^2$ describes a qubit-qubit interaction term. Note that the two terms commute. Initializing the resonator field $\ddd$ in the vacuum, the field state remains in a coherent state and revolves around the qubits-dependent vacuum positions $\alphabold$, as depicted in \cref{fig:chip}b.
After one period of evolution $T = 2\pi/\delta$, the field state comes back to its initial position (the vacuum), and the qubits and the resonator disentangle. The qubits undergo a non trivial evolution $U$ generated by the last term $-\alphabold^2$, with 
\begin{math}U = \exp\left[i\pi(8\bar{n}\chi^2/\delta^2)\JJJ_{\varphi_\Delta}^2\right] \end{math}. When $\delta = 2\sqrt{\bar{n}}\chi$, this implements the entangling gate $U = e^{i\frac{\pi}{2}\JJJ_{\varphi_\Delta}^2}$. With two qubits, $U$ takes the simple form $U = e^{i\frac{\pi}{2}\sss_{zy,1}^{\varphi_\Delta}\sss_{zy,2}^{\varphi_\Delta}}$ up to a global phase factor, where $\sss_{zy,k}^{\varphi_\Delta} = \cos(\varphi_\Delta)\sss_{z_k}-\sin(\varphi_\Delta) \sss_{y_k}$. During the gate, the entanglement of the qubits with the resonator makes the gate fidelity sensitive to the resonator photon loss channel. In addition to $\chi \sqrt{\bar{n}} \ll \Omega_R$, we therefore require that the gate rate is much larger than $\kappa$, i.e $\kappa \ll \chi \sqrt{\bar{n}}$.

\comment{***To see that, one notes that the Hamiltonian~(\cref{eq:simplifiedHamiltonian}) describes a harmonic oscillator driven with a drive of amplitude $-2\sqrt{2\bar{n}}\chi\JJJ_{\varphi_\Delta}$ detuned by $\delta$. Its qubit-dependent equilibrium position is given by $\textbf{x} = -(2\sqrt{2\bar{n}}\chi/\delta)\JJJ_{\varphi_\Delta}$. Initializing the resonator field $\ddd$ in the vacuum, the field state remains in a coherent state and revolves around the displaced vacuum positions $\textbf{x}$. After one period of revolution $T = 2\pi/\delta$, the field state comes back to its initial position (the vacuum), and the qubits and the resonator disentangle. The qubits state has acquired a phase given by the area \begin{math}\textbf{\mathcal{A}}\end{math} in phase space enclosed by the the various paths, with \begin{math}\textbf{\mathcal{A}} = \pi \textbf{x}^2 = \pi(8\bar{n}\chi^2/\delta^2)\JJJ_{\varphi_\Delta}^2 \end{math}. When $\delta = 2\sqrt{\bar{n}}\chi$, this schemes implement the entangling gate $U = e^{i\frac{\pi}{2}\JJJ_{\varphi_\Delta}^2}$.} 

\begin{figure}
% \begin{center}
    \includegraphics[width=\linewidth]{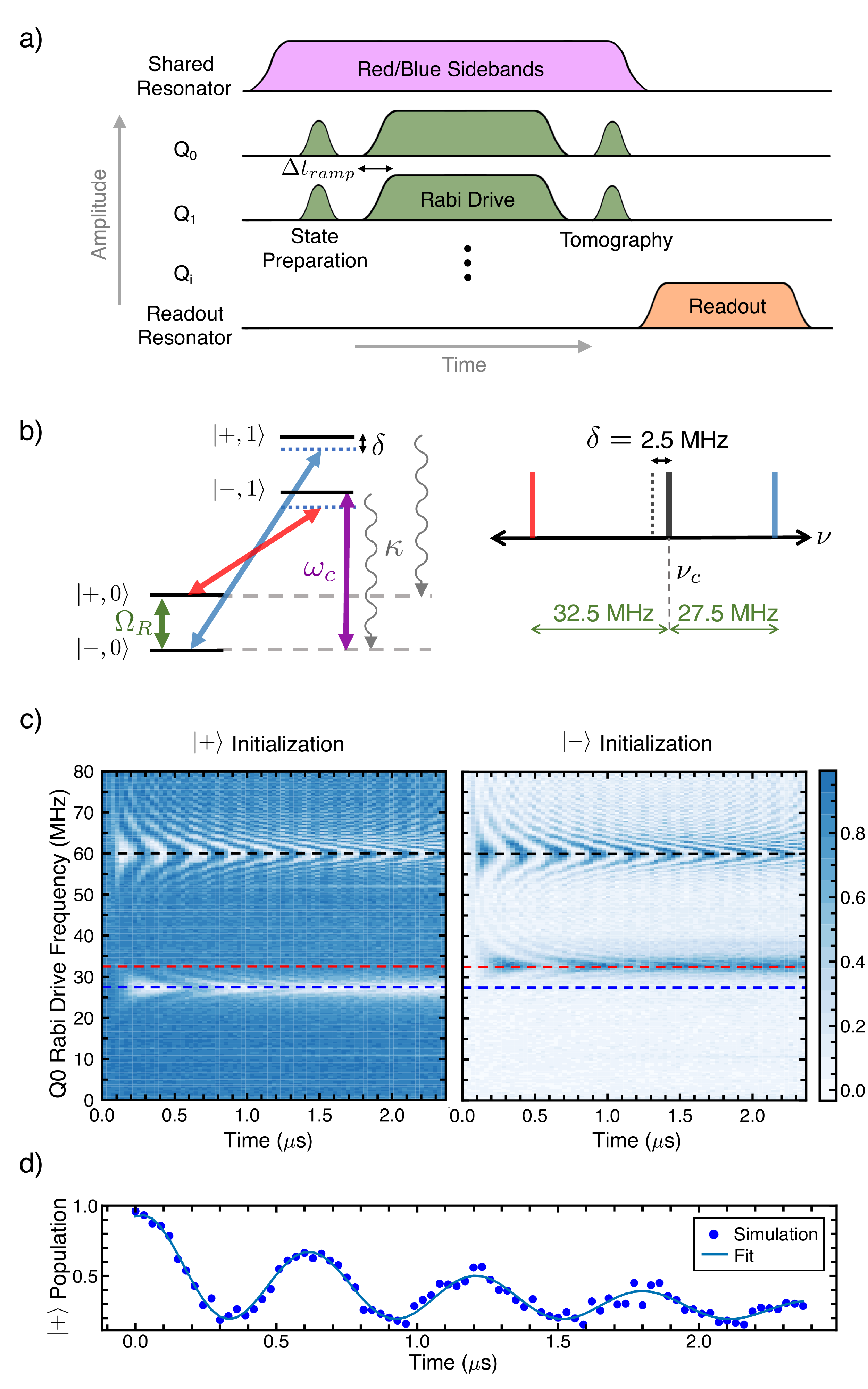}
    \caption{\textbf{Hamiltonian characterization and gate calibration.} a) Pulse sequence for $n$-qubits gate. The sequence for single qubit calibrations and Hamiltonian spectroscopy is identical to the pulse sequence shown with n=1. b) Energy level diagram for a Rabi driven qubit and sidebands. $\kappa$ is the dissipation out of the shared resonator. The choice of qubit initialization pulse determines whether $\kappa$ will either pin the qubit state in $\ket{+}$ or $\ket{-}$. Sidebands are $2\Omega_R = 60$MHz apart with an overall 2.5 MHz detuning down relative the shared resonator frequency. c) Spectroscopy of the Hamiltonian for a single qubit. Spectroscopy is performed by repeating the single-qubit pulse sequence in a) with fixed sideband frequencies while sweeping the Rabi drive frequency, which changes the energy level splittings shown in b). With the sidebands set 60 MHz apart, the gate requires that the qubit Rabi drive be at 30 MHz, which is exactly in between the resonance features indicated by the red and blue dashed lines. d) Cross-section of c) for $\ket{+}$ initialization at the blue dotted line. The continuous blue line is a fit of the data. \label{fig:pulses}}
\end{figure}

To implement the gate, we begin by calibrating the parameters most important to our system: $\chi$, $\bar{n}$, and $\kappa$. The combination of $\chi$ and $\bar{n}$ sets the gate time and is used to calculate the sideband detuning $\delta_{sb}$ necessary for the gate \cite{Murch_2012}. Whereas $\kappa$, the decay rate of photons from the shared resonator, sets a limit on the fidelity of the gate. We discuss this below when we analyse the sources of errors. The pulse sequence shown in \cref{fig:pulses}a for the full gate shows a Rabi drive applied to each participating qubit. We rely on the single Rabi driven qubit version of this pulse sequence to extract all three parameters. We point out that if one sets the state preparation and tomography pulses as $Y_\frac{\pi}{2} = e^{-i\frac{\pi}{4}\sigma_y}$, the gate pulse sequence is the classic spinlocking sequence, done on multiple qubits in parallel under the presence of sidebands \cite{Yan_2013}. We use this protocol as a form of chip and Hamiltonian characterization. 

We provide an overview of our characterization procedure in \cref{fig:pulses}. The level diagram for the single qubit interaction with sidebands is shown in \cref{fig:pulses}b). We apply a state-preparation pulse to initialize the qubit(s) of interest, followed by a Rabi drive around the $x$ axis to bring the qubits close to resonance with the sidebands. For instance in \cref{fig:pulses}c, we sweep $\Omega_R$ with sideband frequencies fixed at $\nu_{res} + \Omega_R - \delta$ and $\nu_{res}-\Omega_R- \delta$ for $\Omega_R = 30$MHz and detuning $\delta = 2.5$ MHz. The qubit is in resonance with the red (blue) sideband when $\Omega_{R}= $ 32.5 (27.5) MHz. When $\Omega_R$ is far from resonance, the dynamics is set by a $T_{1,\rho}$ limited decay of the driven qubit \cite{Yan_2013}, where $T_{1,\rho}$ is the dressed frame analog of $T_1$ qubit relaxation. In contrast, when $\Omega_R$ is close to resonance, excitation swaps will occur in addition to the $T_{1,\rho}$ limited decay. 

We use excitation swaps between the Rabi driven qubit and resonator combined with a Stark shift measurement to calibrate $\chi$ and $\bar{n}$. Where $\chi\bar{n}\gg\kappa$, these exchanges occur at a faster time scale than the loss out of the shared resonator. The red (blue) sideband drives population swaps between $\ket{+,0} (\ket{-,0})$ and $\ket{-,1} (\ket{+,1})$, shown in \cref{fig:pulses}d, which is a cutaway of the full chevron in \cref{fig:pulses}c for $Q_0$. The rate of these oscillations is set by $\chi\sqrt{\bar{n}}$, and the observed exponential decay is set by $\kappa$, cooling the driven qubit to either $\ket{+}$ or $\ket{-}$ \cite{Murch_2012}. This contrasts with previous works studying the resonance features in the low $\chi/\kappa$ that show similar cooling effects without population swaps \cite{Murch_2012,aron_bathEng_entanglement, schwartz_stablizingEnt}. \comment{depending on the initialization of the qubit state and the sideband present. The stabilization effect obtained from bath engineering has been used to generate entanglement.} The frequency of the oscillations in \cref{fig:pulses}d combined with the Stark shift of the qubit - which has a $\chi\bar{n}$ dependence - for different sidebands power allow us to calibrate $\chi$ and $\bar{n}$. We explain this in detail in the Supplement, along with other calibration measurements for the Rabi drive power $\Omega_i$ and the sideband phase. We note that there is an additional resonance feature at 60 MHz, marked by the black dotted line in \cref{fig:pulses}c. These are generated by higher order terms in the Hamiltonian due to the presence of the Rabi drive. In general, as discussed later, $\omega_b - \omega_r$ should be maximized as to push this higher frequency resonance away from the Rabi drive frequencies for the gate. 

%However, compared to a directly tunable qubit we have an added advantage that the Rabi drive decouples the qubit from noise and increases relevant lifetimes. Instead of being limited by T1 and T2 lifetimes, we are now limited by $T_{1,\rho}$ and $T_{2,\rho}$ lifetimes. At the Rabi drive used to implement the two-qubit gate (30MHz), $T_{1,\rho} =\mu$s and $T_{2,\rho} =\mu$s for Q1 [Q2] and the single qubit fidelities measured from randomized benchmarking are ..... and ..... (see Ref. [...]) for benchmarking data and coherence measurement
\section{Gate Characterization} \label{sec:sec3}

\begin{figure}[ht]
    \includegraphics[width=\linewidth]{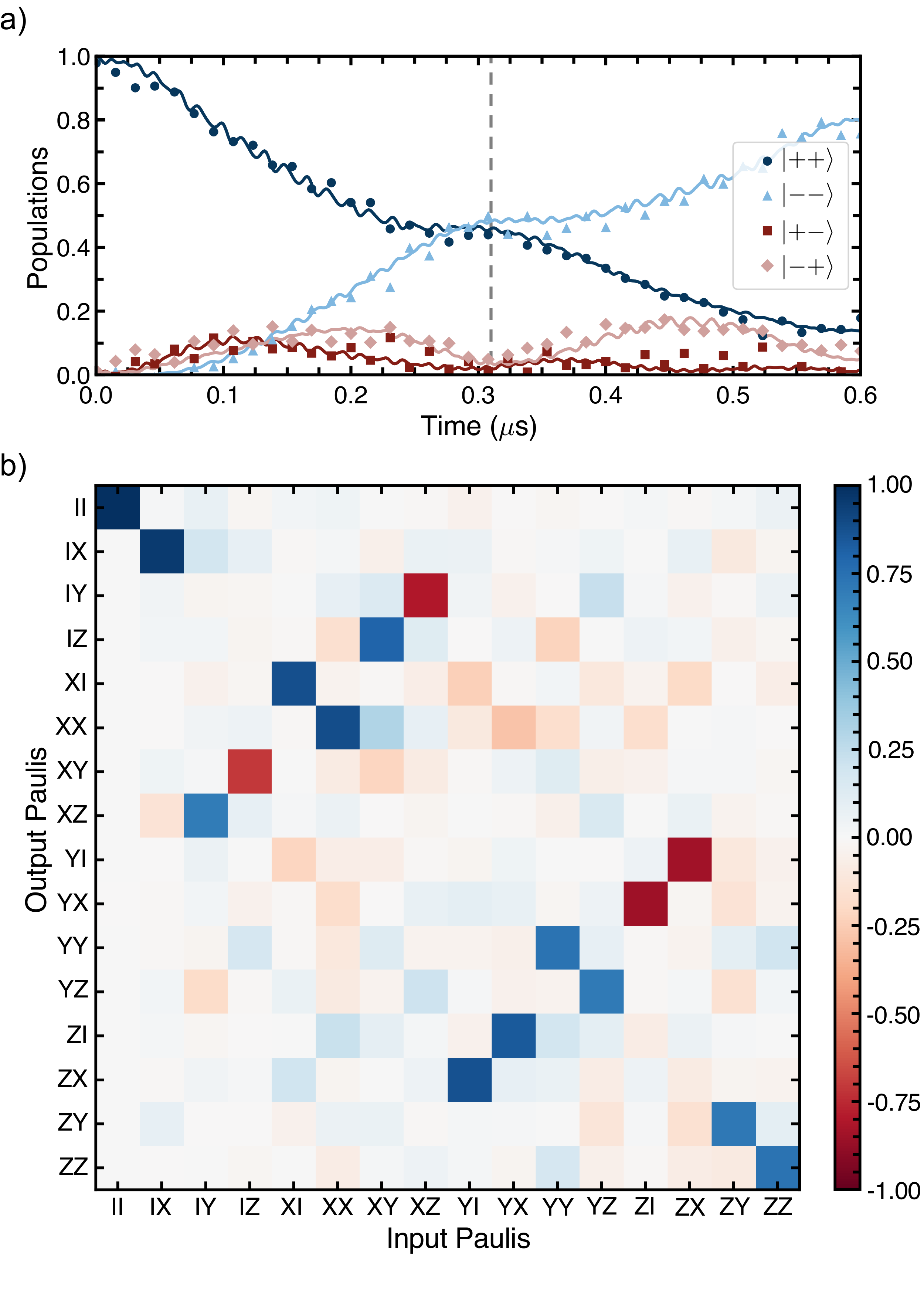}
    \caption{\textbf{Characterizing two-qubit gate.} a) Gate evolution for two qubits. Experimental data (points) together with simulations of \cref{eq:masterEq} (solid lines) with $\Omega_0/2\pi=30.55$ MHz, $\Omega_1/2\pi=29.92$ MHz, $\chi_0/2\pi = 380 KHz$, $\chi_1/2\pi=410$ KHz, $\kappa=180$ KHz. b) PTM of the experimental data in the $\ket{g/e}$ basis, at $T_{gate} = 310$ ns, obtained measuring 16 different initial states. The process fidelity is 81.6$\%$. The average fidelity to the target bell states 91.8\%. The solid lines that fit each population are from simulations. \label{fig:ptm}}
    % \caption{Gate evolution for 2 qubits. a) Simulations without decoherence for a Rabi drive of 30 MHz, with initial state $\ket{-,-}$. The 2 driven qubits entangle and disentangle with the shared resonator with time, with full disentanglement at the gate time of 0.24 $\mu$s, shown by the dashed vertical line. The maximum concurrence and fidelities exceed 99.9$\%$ at the gate time. b) Experimental data together with simulations with $\Omega_1=29.8$ MHz, $\Omega_2=30.1$ MHz (\red{simus numbers, change and would be better to have a true fit. Recheck which kappa I used, I think it was low.}), including decoherence, leading to a maximum fidelity of 93.0$\%$. c) PTM of the experimental data in the g,e basis, at $T_{gate} = 310$ ns, obtained measuring 16 different initial states. The process fidelity is 92.0$\%$.}
\end{figure}

%(https://arxiv.org/pdf/1202.5344.pdf)

Next, we study the two-qubit population evolution over time to extract the gate time. To perform the gate, we implement the full pulse sequence in \cref{fig:pulses}a on any desired subset of our qubits. We use $Q_0$ and $Q_1$, two qubits with the most similar shared resonator dispersive couplings $\chi$. The results for our choice of $\ket{++}$ initial state are summarized in \cref{fig:ptm}. Maximum entanglement occurs when the $\ket{+-}$ and $\ket{-+}$ populations reach a minimum and the $\ket{++}$ and $\ket{--}$ populations cross, as indicated by the vertical dashed line in \cref{fig:ptm}a. The gate time at that point is 310 ns, which refers to the length of the Rabi drive applied to each qubit. We prepare each of the 4 Bell states and perform state tomography at that gate time. The state fidelity \cite{nielsenChuang}, defined as \begin{equation} F = \sqrt{\sqrt{\rho}\sigma\sqrt{\rho}},\end{equation} ranges from 87\% to 95\% with an average of 91\%.

We further characterize our gate using quantum process tomography (QPT) \cite{chuang_QPT, poyatos_QPT}, which is achieved by preparing 16 different input states and performing state tomography on each output state. A convenient set of input states that spans the two-qubit subspace is $\{\ket{+}, \ket{-}, \ket{g}, \ket{i-}\}\otimes \{\ket{+}, \ket{-}, \ket{g}, \ket{i-}\}$, where $\ket{i-} = (\ket{+}-i\ket{-})/\sqrt{2}$. The measurement basis is chosen as $\sigma_m \otimes \sigma_n$ where $m,n = {0, x, y, z}$, and $\sigma_0 = I$. Since we measure the qubits in the $\sigma_z$ basis, we apply single qubit tomographic pulses to measure the expectation values of $\sigma_i\otimes\sigma_j$. We present our results as a Pauli Transfer Matrix in \cref{fig:ptm}b, which maps input Pauli state vectors to output Pauli state vectors. We then perform QPT on the two-qubit entangling interaction with process fidelity of 81.6\% between qubits 0 and 1. The average state fidelity of all 16 initial states is 91.8\%. We note that the qubit Rabi drive adds a global phase in the Rabi driven qubit frame that we unwind using techniques described in the Supplement, resulting in the PTM shown. We also perform QPT on the identity operation with process fidelity 93.5\% as a baseline check of state preparation and measurement (SPAM) errors (see \cref{fig:ptm_id} in Supplements). 

\begin{figure}[t!]
    \includegraphics[width=\linewidth]{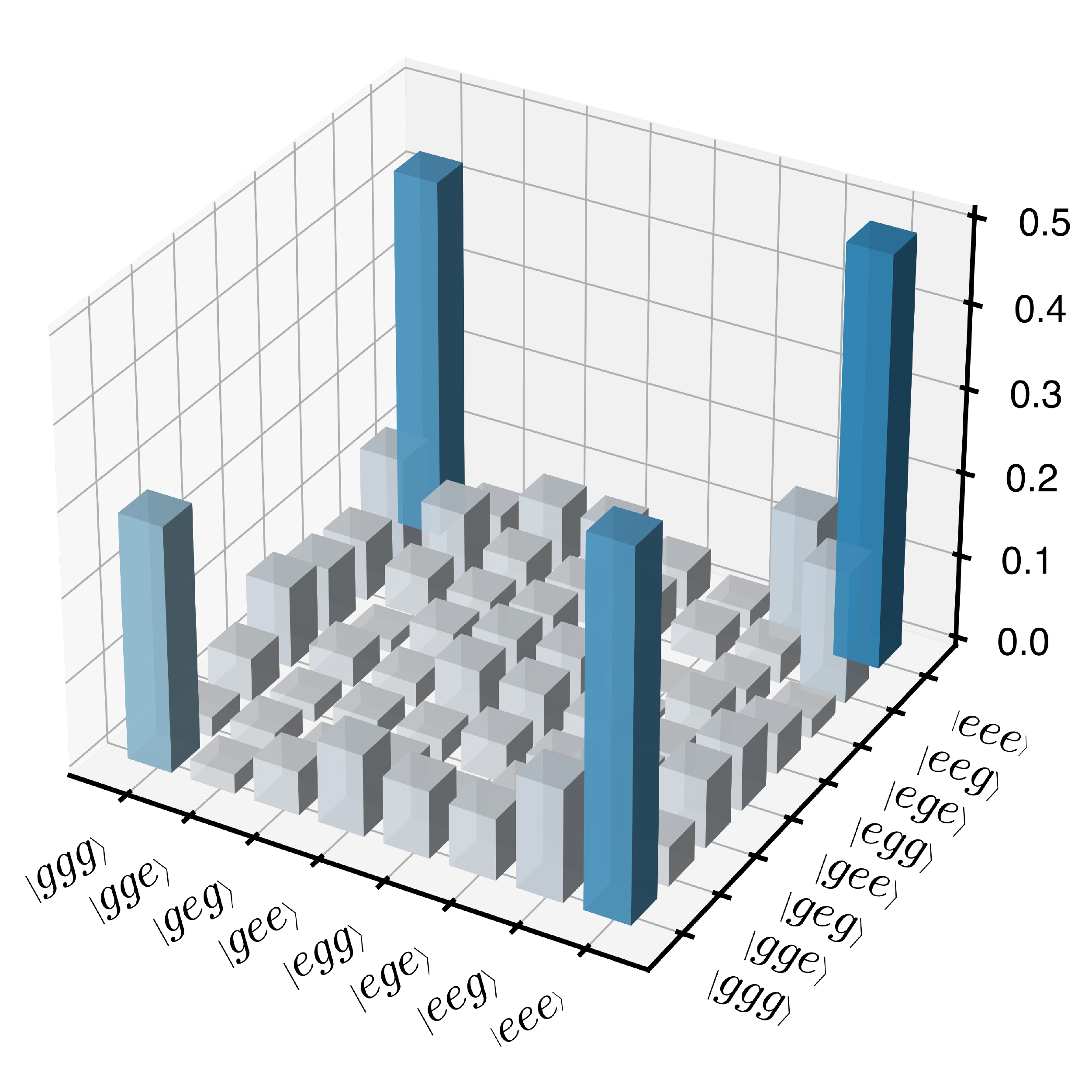}
    \caption{\textbf{Three-qubit entanglement.} Density matrix for three qubits measured at gate time of 217 ns. The bars represent the amplitude of the density matrix elements. We prepare the qubits in $\ket{+++}$ state and use sideband phase calibrated to be 180 degrees offset from the value used for the two-qubit interactions such that we entangle directly in the bare qubit basis. Note here that the qubit states are labeled with $\ket{g/e}$ rather than $\ket{\pm}$. Entanglement in the $\pm$ basis can be done by preparing in the $\ket{i-}$ state and using the same sideband phase as in the 2 qubit measurements. We chose this for convenience.   \label{fig:3Q}
    }
\end{figure}

The most important feature of the gate is its scalability, which we demonstrate by generating a three-qubit GHZ state on $Q_0, Q_1, Q_2$: \begin{equation} \Psi_{\text{GHZ}} = \frac{\ket{ggg} + \ket{eee}}{\sqrt{2}}.\end{equation} We note here that for odd numbers of qubits, the Hamiltonian allows us to generate entanglement in either the $\ket{\pm}$ basis or directly in the $\ket{g/e}$ basis, depending on the initial state and sideband phase. Given that the two bases have a direct mapping between them, we choose an initial state and sideband phase combination that directly entangles in the $\ket{g/e}$ basis for simplicity. We use a $Y_{\frac{\pi}{2}}$ pulse to prepare each qubit in the $\ket{+}$ state and then apply a $\sigma_x$ Rabi drive to each qubit for 217 ns. Using individual qubit pulses to do state tomography, we find state fidelity 90.5\%, similar to our two qubit fidelities. The density matrix is shown in \cref{fig:3Q}. We note that the three-qubit gate time is faster than the two-qubit gate time because the speed is proportional to the average $\chi_i$ and $\bar{n}$. We used the same sideband power for both gates, but the third added qubit had a much higher coupling than the first two qubits, thus raising the gate speed. 

Finally, we attempted 4-qubit GHZ state preparation and achieved state fidelity 66\% in 200 ns. The density matrix is shown in the Supplement \cref{fig:4Q_dms}. Similar to before, the gate time is even shorter because the final added qubit has the highest $\chi$ of all qubits on our processor. The 4-qubit fidelity is mainly limited by the large spread in $\chi$ and crosstalk between $Q_2$ and $Q_3$. The $Q_3$ qubit frequency is very close to the $e \longrightarrow f$ transition frequency for $Q_2$. When both qubits are Rabi driven, as during the gate, we see significant $f$ state population for $Q_2$. To mitigate this, we had to lower $\Omega_R$ for each qubit to 20 MHz, which further limits fidelity. We discuss this in our error analysis. 
\section{Discussion and Errors} \label{sec:sec5}

\begin{figure}
    \includegraphics[width=\linewidth]{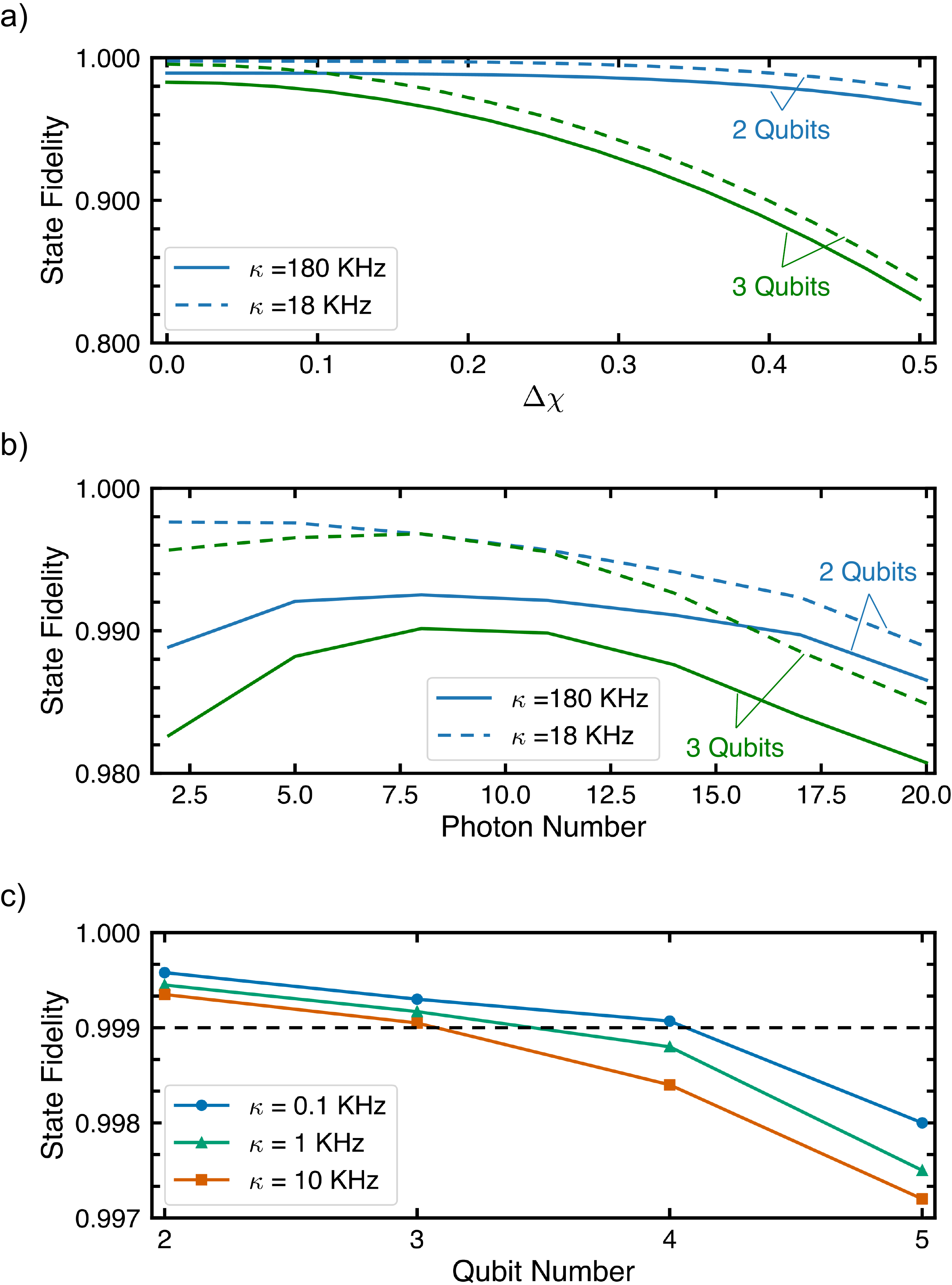}
    \caption{\textbf{Error analysis using simulations of \cref{eq:masterEq}} a) For a system composed of two (blue) or three (green) qubits, the fidelity of the gate is shown as a function of $\Delta \chi$. The two different styles of lines represent different $\kappa$ values. We exclude qubit in order to isolate the contribution of error due to $\Delta\chi$ and use the average of the $\chi_\text{av}/2\pi = 500~\text{KHz}$.  b) State fidelity as a function of the mean photon number $\bar{n}$. A higher photon number raises the gate speed which helps mitigate the effect of lifetimes but also raises the contribution of counter rotating terms that cause the fidelity to saturate at approximately 10 photons. c) Scaling of entanglement fidelity as function of N, the number of qubits, for the best possible achievable chip. The black dashed line marks the 0.999 fidelity threshold. We set $\Omega_R=$150 MHz and $\chi_k = 1$MHz while varying $\kappa$. All three figures take qubit lifetimes to the industry best values.  \label{fig:theory_prediction}
    }
\end{figure}
%For a given ratio $\Omega/\kappa$, the absolute values of $\Omega$ and $\kappa$ are chosen such that the infidelity due to losses equals the infidelity due to finite ratio of $\Omega/\chi \bar{n}$.
%The parameters are set to $\kappa/2\pi = 2~\textrm{kHz}$, $T_{1\rho}=T_{2\rho} = 400~\textrm{us}$, $\chi_\text{av}/2\pi = 0.4~\text{MHz}$, and  $\Delta\chi = 10\%$. The black dashed-dotted line corresponds to a simulation where the spurious term oscillating at $2\Omega$ is dropped.
We now turn to a study of the following sources of infidelity: state preparation and measurement (SPAM) errors, shared resonator decay ($\kappa$), spread in qubit-shared resonator couplings ($\Delta\chi$), $\delta$ calibration errors, pulse shaping, and the effects of a lower Rabi drive rate ($\Omega_R$). Each error listed here is included in our simulations aside from SPAM errors. 

The first major source of infidelity is due to state preparation and measurement (SPAM) error. Process tomography of the identity process resulted in process fidelity of 93.5\%. Due to the nonzero $\kappa$ of the shared resonator, we start the sideband pulse before the state preparation pulse and perform all qubit operations, including the tomography pulses, under the presence of sidebands. However, the presence of sidebands decreases our qubit lifetimes and pulse quality (see Supplement). 

Assuming no SPAM errors, simulations based on \cref{eq:masterEq} (details found in the Supplementary Materials) suggest $\kappa$ as leading sources of error (a complete summary of simulated accumulated error can be found in \cref{table_lifetimes}). The effect of $\kappa$ is shown in the decay of oscillations in \cref{fig:pulses}d. During the course of the gate, the resonator makes a circle in phase space (\cref{fig:chip}c) and is populated with maximal photon population $\bar{n}=0.5$ as the qubits exchange excitations with the resonator. This allows the qubits to each simultaneously entangle and then disentangle with the resonator, leaving the qubits entangled with one other. However, while the resonator is populated, photons may decay from the resonator. We have designed the external $\kappa$ of the resonators to minimize the loss of photons during the gate. As such, the resonator loss is dominated by internal loss. \comment{Simulations predict that for  $\kappa = 380  \text{kHz}$ -- as in our system -- we expect 1.7\% (1.9\%) gate infidelity for two (three) qubits. Where as qubit lifetimes alone only contribute an additional 0.6\% (0.7\%).}

A second source of error that becomes more important with increasing qubit number is the spread of cross-Kerr couplings, $\Delta\chi$, between the qubits. We define $\Delta\chi=\frac{\text{max}(\chi)-\text{min}(\chi)}{\text{avg}(\chi)}$. Our simulations show that for a two qubit gate, the contribution of $\Delta\chi$ to the infidelity is overshadowed by $\kappa$. However, for a three-qubit gate, $\Delta\chi$ carries a greater weight. To verify this experimentally, we measure the two-qubit state fidelity between two qubits ($Q_0$, $Q_2$) that have the largest $\Delta\chi$ among the three qubits used for our multi-qubit gate. We obtain a state fidelity of 93\% for preparing $\frac{\ket{++} + \ket{--}}{\sqrt{2}}$ , which is very similar to maximum fidelities observed between $Q_0$ and $Q_1$. However, while the three-qubit gate uses the same qubits as in our two-qubit experiments, we see a drop in state fidelity as $\Delta\chi$ as a larger effect for increasing numbers of qubits. This takes even greater effect for the four-qubit gate, as $Q_2$ and $Q_3$ have almost twice the coupling strength as $Q_0$ and $Q_1$. We note that while the fidelity does strongly depend on $\Delta\chi$, the couplings on this chip were anomalies due to design and fabrication errors and standard fabrication techniques should allow for $\Delta\chi$ values below 15\%.

To further describe the effects of $\kappa$ and $\Delta\chi$, we again performed simulations with \cref{eq:masterEq} as a function of the parameters of interest.
We summarize the effects of $\kappa$ and $\Delta\chi$ in \cref{fig:theory_prediction}. In \cref{fig:theory_prediction}a and b, we use resonator loss values that are similar that on our sample and show fidelity dependence on key system parameters. We take qubit lifetimes to be infinite in order to isolate the effect of each parameter. \cref{fig:theory_prediction}a shows the fidelity as a function of $\Delta\chi$, $\kappa$, and the number of qubits. \cref{fig:theory_prediction}b shows expected fidelites as a function of the sideband power. While increasing $\bar{n}$ raises fidelites at first because it decreases the gate time compared to loss rates, increased $\bar{n}$ also widens the $2\Omega_R$ feature shown in \cref{fig:pulses}c, increasing the contribution of spurious counter-rotating terms in the Hamiltonian. Our simulations (\cref{fig:theory_prediction}c) suggest that for leading fabrication techniques that produce resonators with quality factors over 5 million and leading qubit coherence times \cite{millionQ,best_coherence1}, the gate can be implemented at or above the fault tolerant threshold of 0.999 fidelity for up to four qubits. Furthermore, the STAR gate can be useful for running algorithms on NISQ processors for even higher numbers of qubits. 

While $\kappa$ and $\Delta\chi$ are the main error sources that we are able to characterize experimentally, there are a few additional contributions that will be important to consider to optimize for the best attainable fidelities. First, we are currently not verifying how well the qubits disentangle from the shared resonator at the gate time. Ideally, a proper calibration of the $\delta$--obtained from our measurements of $\chi$ and $\bar{n}$--should ensure this, but because we do not measure the state of the resonator, there is the possibility of an inaccurate calibration. As an example, from simulations, a miscalibration of $\delta$ by 10\% on a two-qubit gate would add an additional 1.3\% error. 
Second is the finite ramp time used in the shape of the Rabi pulse. This ramp is necessary to keep the spectrum of the pulse narrow in frequency space. However, as the drive ramps up to the required Rabi frequency, it crosses a resonance with the sidebands. These are the same resonances used in the spinlocking measurements to calibrate our chip parameters in \cref{fig:pulses}. These spurious interactions can be mitigated using pulse optimization techniques \cite{pulse_shaping1, pulses_shaping2}.

The final source of error is Rabi drive rate. As shown in the chevron plot in \cref{fig:pulses}c, there is a feature at frequency $2\Omega$ marked by the black dotted line, at twice the Rabi drive frequency typically used for the gate at $\Omega_R$. The frequency of the oscillations of the $2\Omega_R$ feature are at $2\times 30 \text{MHz} = 60 \text{ MHz}$, and the width of the oscillations is set by $\chi\bar{n}$. For higher sideband powers and lower Rabi drive frequencies, the oscillations generate counter-rotating terms in the Hamiltonian that interfere with gate dynamics. In our numerical simulations, the effects on the gate can be seen in \cref{fig:sim_30vs60} in the Supplement (comparing 30 and 60 MHz Rabi drive gates). The fidelity of the two-qubit gate would benefit from increasing the Rabi drive to over 100 MHz, or equivalently, adding an extra cancellation tone to offset the effects of the spurious feature.

\section{Conclusions and Outlook} \label{sec:conclusions}
    In summary, we demonstrate a scalable maximally entangling gate on an all-to-all connected fixed frequency transmon processor between two, three, and four qubits. The gate is generated through bichromatic microwaves and Rabi drives applied to each participating qubit. The Rabi drive provides the advantages of reducing limitations on qubit-qubit detunings during fabrication, dynamically decoupling from noise, and allowing us to entangle any subset of qubits on the chip. Finally, for the three-qubit gate, we are able to choose whether to entangle the qubits in the Rabi dressed basis or the original qubit basis based on the sideband phase and preparation state, without the need for additional qubit pulses to map between bases. Looking forward, the gate is most limited by photon loss from the shared resonator for four and more qubits. Applying state of the art fabrication techniques will yield multi-partite gates that exceed the 0.999 fidelity threshold for up to four qubits. At the same time, for gates with higher numbers of qubits, it is also worth exploring this scheme on parametrically coupled chips where can perform the gate in a far-detuned regime, thus eliminating the $\kappa$ loss factor. 
\section*{Acknowledgements} \label{sec:acknowledgements}
We would like to thank Lucas Buchmann, Felix Motzoi, Kyunghoon Lee, Ravi Naik, Bradley Mitchell, for valuable discussions. This work was undertaken in part thanks to funding from NSERC, the Canada First Research Excellence Fund, the Minist\`ere de l'\'Economie et de l'Innovation du Qu\'ebec, the U.S. Army Research Office under Grant No. W911NF-18-1-0411, the U.S. Department of Energy, Office of Science, National Quantum Information Science Research Centers and Quantum Systems Accelerator, and the L'Or\'eal USA For Women in Science Fellowship Program.

\bibliographystyle{apsrev4-1}
\bibliography{references}
\newpage
\appendix*
\section{Supplementary Material} \label{sec:appendix: Calibrations}
%The coupling strength of sidebands to the Floquet qubit is calibrated by recording resonant Rabi oscillations on the Floquet qubit transition. We perform spinlocking measurements developed by the NMR community (see Ref. ...) ) under the presence of sidebands. The pulse sequence is shown in \cref{fig:pulses} a), where we only use one qubit. The state preparation pulse used is $\pm\pi/2$ to initialize in the $\ket{+}$ or $\ket{–}$ state and the rabi drive lifts the degeneracy between the two states. The energy splitting is proportional to the strength of the drive.  The red (blue) sidebands are resonant with the $\ket{+,0}$ ($\ket{-,0}$) and $\ket{-,1}$($\ket{+,1}$) transitions as shown in \cref{fig:pulses} b). The resulting populations swaps are used to calibrate both chip and gate parameters.
\comment{In \cref{table_chip_params}, we list some relevant frequencies. We point out that the two qubit gate was done on Q0 and Q1, while the 3-qubit gate used Q0, Q1, and Q2. Q2 has much stronger coupling relative to Q0 and Q1, which we believe to be one of the limitations on the gate. 
\begin{table}
    \begin{tabular}{ |c|c| } 
    \hline
    Parameter & value\\
    \hline
     $\omega_{ge}^0$ & 5.24 GHz\\ 
     $\omega_{ge}^1$ & 5.37 GHz\\
     $\omega_{ge}^2$ & 5.69 GHz\\
     $\omega_{ge}^3$ & 5.48 GHz\\
     $\chi_0$ & 380 KHz\\
     $\chi_1$ & 410 KHz\\
     $\chi_2$ & 718 KHz\\
     $\chi_3$ & 815 KHz\\
     $\omega_c$ & 7.82 GHz\\
     $\kappa_{ext}$ & 20 KHz \\
     $\kappa_{int}$ & 180 KHz\\
    \hline
    \end{tabular}
    \caption{Summary of commonly referenced chip parameters}
    \label{table_chip_params}
\end{table}
}

\begin{table}
    \centering
    \begin{tabular}{ |p{0.25\linewidth}|p{0.33\linewidth}|p{0.33\linewidth}| } 
    \hline
    Error Term & Infidelity (2-qubits) & Infidelity (3-qubits)\\
    \hline
     $\Omega = 30$ MHz & 0.14\% & 0.27\%\\
     $\kappa = 180$ KHz & 1.7\% & 1.9\%\\
     $\Delta\chi$ & 2.14\% & 6.8\% \\
     Qubit lifetimes & 2.7\% & 7.5\% \\
     $\delta_{sb}$ miscalibration by 10\% & 4\% & 11\% \\

    \hline
    \end{tabular}
    \caption{Accumulated infidelity with each added term, descending, using parameter values that are currently found on our chip. }
    \label{table_errors}
\end{table}

\subsection{Calibration of $\chi$ and $\bar{n}$.}
\begin{figure}
    \begin{center}
        \includegraphics[width=\linewidth]{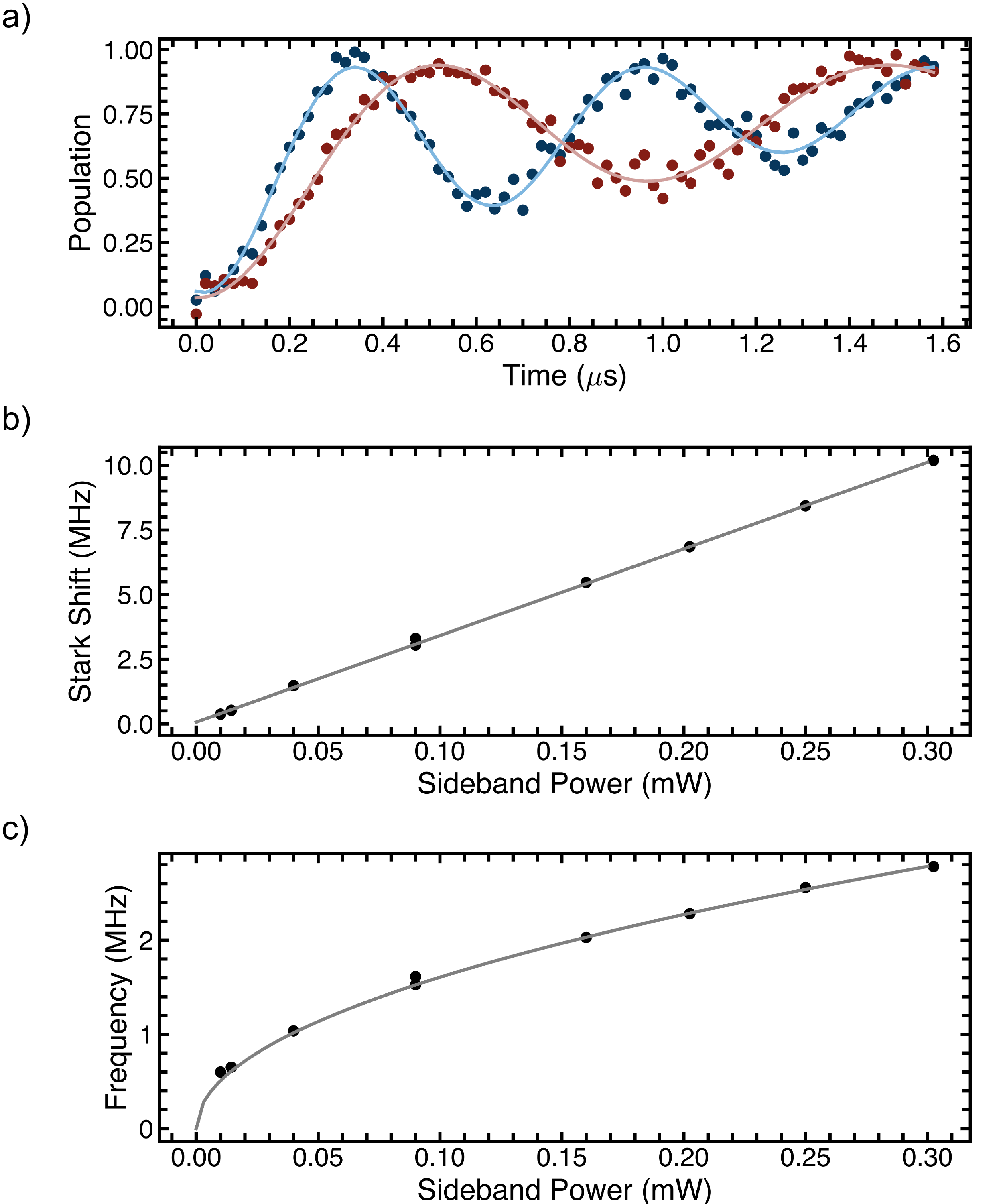}
        \caption{Calibration of $\chi$ for Qubit 3. a) Results of the spinlocking measurement with sidebands on. Oscillations between $\ket{+}$ and $\ket{-}$ states observed for different powers of the sideband. We show two examples here at two different sideband strengths. b) Stark shift vs power of sidebands. c) Frequency of the fitted oscillations from a) vs power of sidebands.}
        \label{fig:SShift_Murch}
    \end{center}
\end{figure}

	We use the combination of two measurements to calibrate the coupling strength of the qubit to sidebands: the spinlocking measurements with sidebands and qubit Stark Shift measurements as a function of sideband power. Both measurements produce oscillations that we fit to extract the frequency. The resulting frequencies for the spinlocking and starkshift measurements exhibit different scalings with respect sidebands power. In the spinlocking sequence, we repeat the measurement for several different red sideband powers, recording the resonant population swaps between $\ket{+0}$ and $\ket{-1}$. We fit the population evolution and extract the frequency of the oscillations, which has a $\chi \sqrt{\bar n}$ dependency, and sets the strength of the couplings between the floquet qubit and sidebands, as shown in \cref{fig:SShift_Murch}. This $\chi \sqrt{\bar n}$ value sets the gate detuning and also gate time. We also measure the Stark shift of the bare qubit frequency as a function of the sideband strength using a Ramsey measurement, for which we expect a $\chi \bar n$ dependence. Combining these two relations, we obtain the $\chi$'s. 

\begin{figure}
    \begin{center}
        \includegraphics[width=\linewidth]{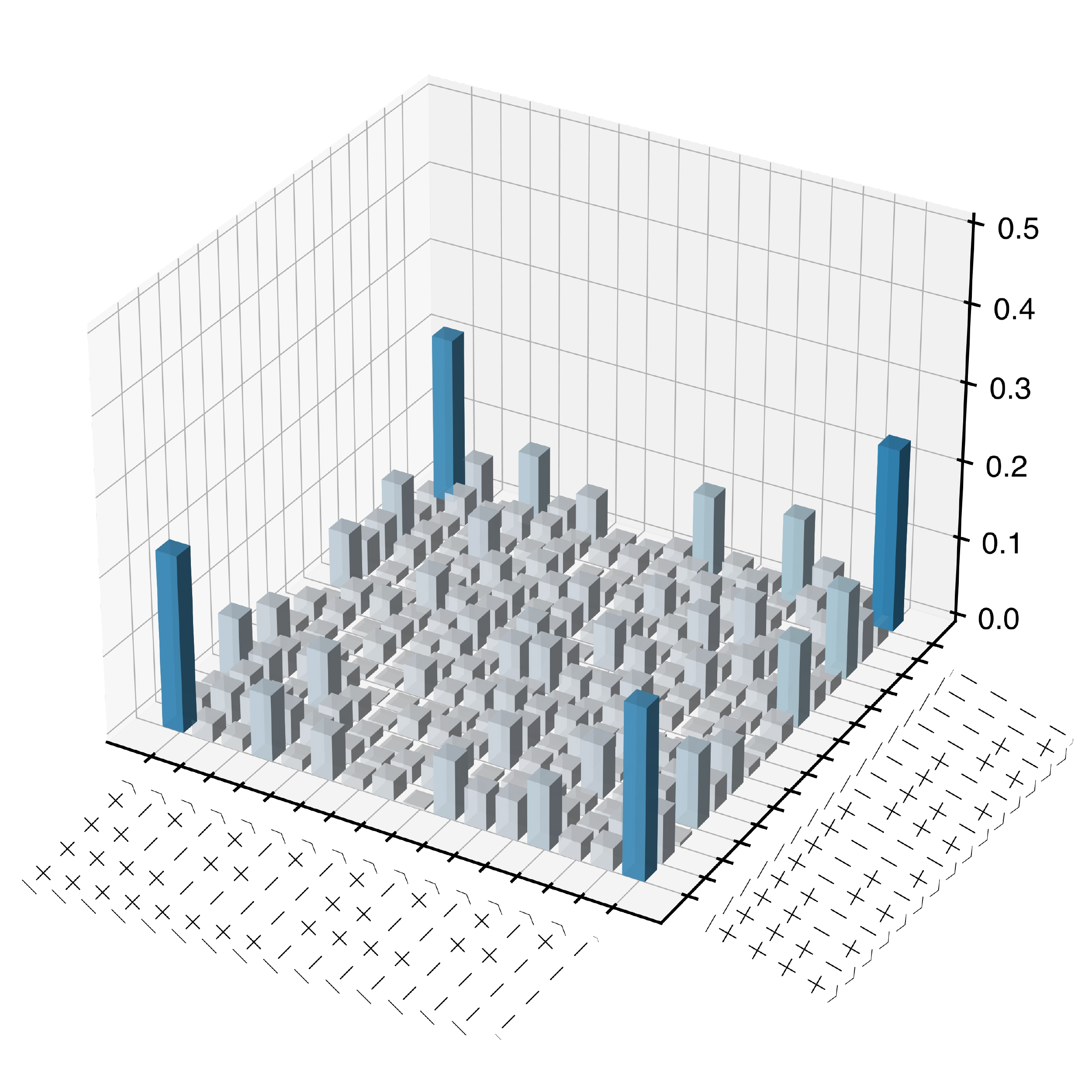}
        \caption{4Q density matrix}
        \label{fig:4Q_dms}
    \end{center}
\end{figure}
\begin{figure}
    \begin{center}
        \includegraphics[width=\linewidth]{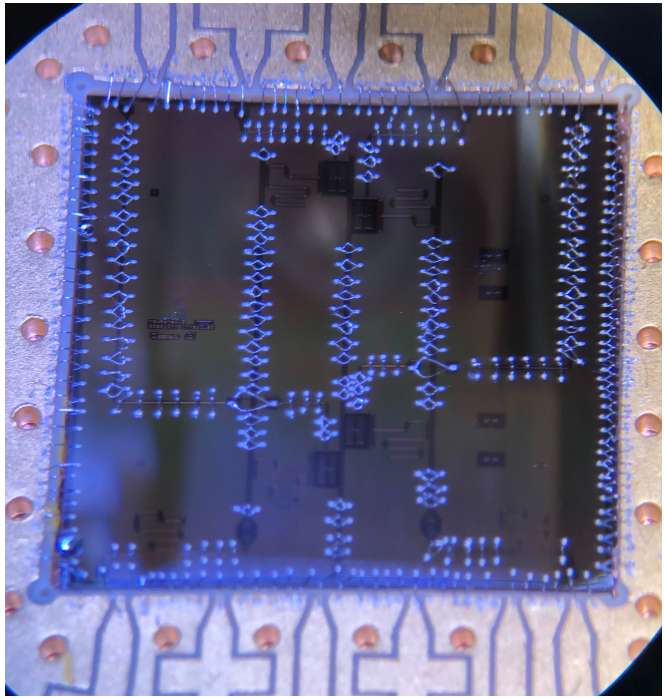}
        \caption{Chip used for experiment.}
        \label{fig:chip_exp}
    \end{center}
\end{figure}

\subsection{Rabi Drive}
\begin{figure}
    \begin{center}
        \includegraphics[width=\linewidth]{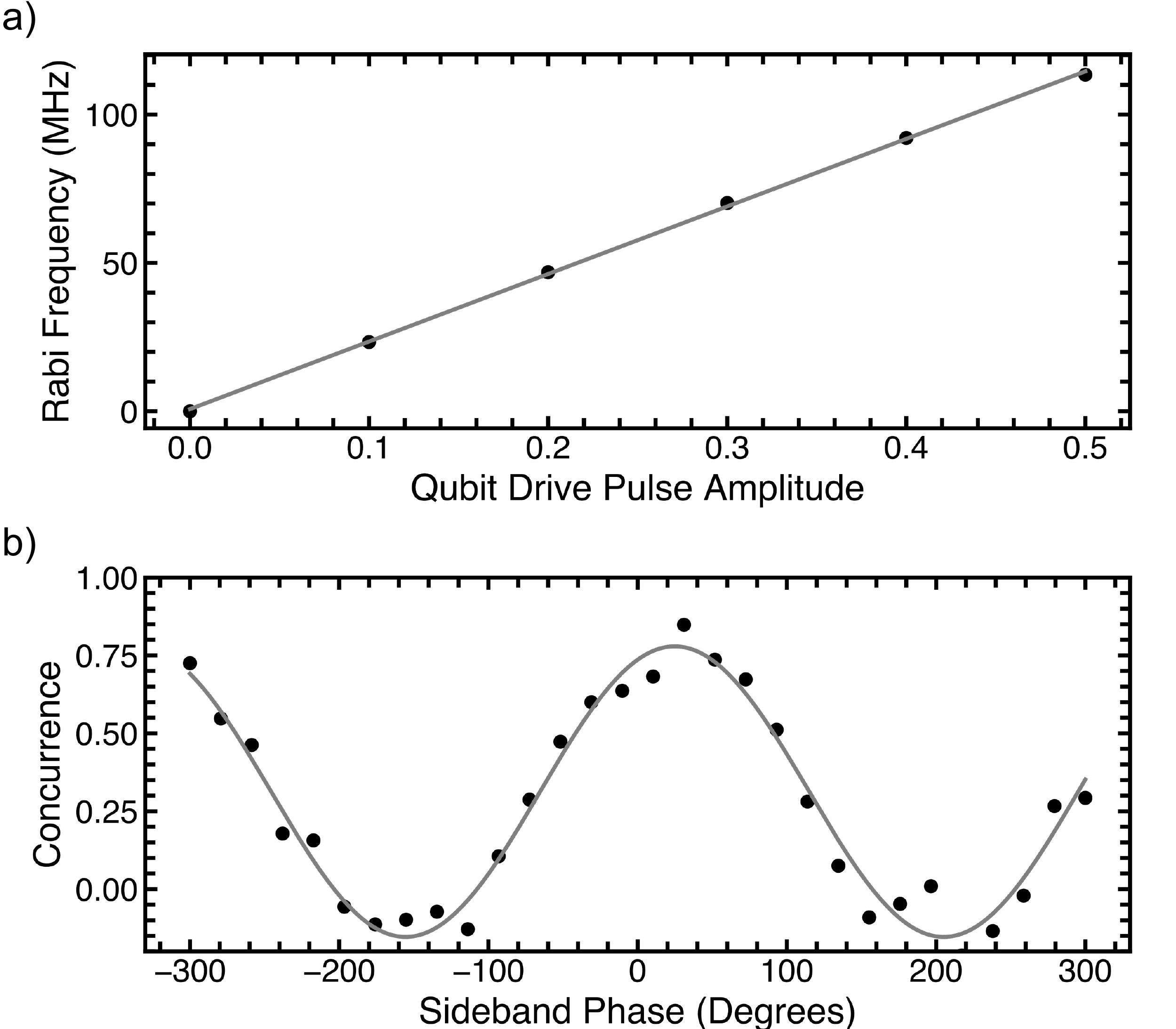}
        \caption{a) Fitted Rabi frequencies as a function of the amplitude of the qubit pulse used to do the Rabi drive. b) Sideband phase calibration}
        \label{fig:rabiSBCalib}
    \end{center}
\end{figure}

	We sweep the amplitude of the qubit pulse for repeated Rabi drive measurements. The population swaps between $\ket{g}$ and $\ket{e}$ are fit with a $\sin()$ function and the frequency is extracted. We then plot the frequencies as a function of the amplitude of the pulse, producing a linear dependence (\cref{fig:rabiSBCalib}a). The linear fit is used to interpolate between the data points to select the amplitudes required for specific Rabi drive frequencies. 

\subsection{Sidebands Phase calibration}
	The phase difference of the sidebands and their relative phase compared to the qubit drive determines whether the Hamiltonian is an XX or combination of XX and YY with respect to the $\ket{\pm}$ basis. Due to the fact that the Rabi pulse we use for the gate has a cosine edge, there an extra accumulation of phase difference between the sidebands and the qubit drive depending on the length of the cosine edge. We calibrate the sideband phase such that we implement the XX interaction. One can tune the gate angle $\varphi_\Delta^\text{eff} = \varphi_\Delta (t_r)-\Omega_R t_r =\varphi_\Delta (0)+\Omega_R t_r$ by adjusting the sideband phase difference $\varphi_\Delta (0)$. The angle is calibrated by initializing the qubits in $\ket{00}$, running the gate and looking at the concurrence as a function of the sideband phase. The effective angle $\varphi_\Delta^\text{eff}$ is set to 0 when the concurrence is minimum. In Fig \cref{fig:rabiSBCalib}b

\subsection{Unwinding Rabi drives of the gate}

The gates applied on undriven transmon qubits commonly take place in the frame rotating at the mode frequencies. In the case of Rabi driven qubits, the system undergoes the MS gate in the frame rotating at the Rabi frequencies.
As the latter is not kept constant throughout the entire pulse sequence and as the tomography is done in the original undriven frame, one needs to track and unwind the phase accumulated to characterize the gate. 

The Rabi drive pulse, of duration $t_p$, consists a square amplitude pulse, of duration $t_{sq}$ during which the entangling operation is happening, sandwiched between two cos edge ramps, of duration $t_{r}$. The gate occurs during the square pulse, and one has $t_p = t_{sq} +2 t_r$.

Neglecting decoherence, the propagator corresponding to the full Rabi drive pulse can be expanded as $U_{t_p,0} = U_\text{down}U_\text{gate} U_\text{up}$,
with $U_\text{down} = U_{t_{sq}+2t_r,t_{sq}+t_r}$, $U_\text{gate}(t_{sq}) = U_{t_{sq}+t_r,t_r}$, and $U_\text{up} = U_{t_r,0}$. Note that the populations in \cref{fig:ptm} are plotted as a function of $t_{sq}$.

As the ramp time  $t_r$ is much shorter than the inverse gate rate, the qubit-resonator coupling is (almost) resonant only during the square pulse. Consequently, $U_\text{down}$ and $U_\text{up}$ consists of local operations on each qubit around the X axis (accumulated phase), given by $U_\text{up/down} = \exp(-i\Omega_R t_r \JJJ_x/2)$, where $\JJJ_x = (\sss_{x_1}+\sss_{x_2})/2$. The factor $1/2$ in $U_\text{up/down}$ comes from the integral of the cos edge pulse. Here, we have assumed that the Rabi frequencies of the qubit are the same.

The propagator $U_\text{gate}(t_{sq})$ expands as $U_\text{gate}(t_{sq}) = U_{R}(t_{sq}) U_\text{gate}^{R}(t_{sq})$, where $U_{R}(t_{sq}) = \exp(-i\Omega_R t_{sq} \JJJ_x)$ represents the single qubit phase accumulated during the gate, and $U_\text{R,gate}(t_{sq}) = U_{t_{sq}+t_r,t_r}^R$ is the time evolution operator generated by the Hamiltonian in \cref{eq:simplifiedHamiltonian}. Since the two sideband tones are detuned from one another by $2\Omega_R$, the effective sideband phase difference $\varphi_\Delta$ of the gate depends on the time $t_r$ at which the gate starts. More precisely, one can write $\varphi_\Delta (t_r) = \varphi_\Delta (0)+2\Omega t_r$.

The global evolution operator can be cast in the convenient form 
$U_{t_p,0} = U_\text{down} U_{R}(t_{sq})  U_\text{up} (U_\text{up}^\dagger U_\text{R,gate}^{\varphi_\Delta (t_r)}(t_{sq}) U_\text{up})$. From the expression of $U_\text{up/down}$, we note that this operator applies a rotation on the qubits around the $\JJJ_x$ axis by an angle $\Omega_R t_r$, and its effect on $U_\text{R,gate}^{\varphi_\Delta (t_r)}(t_{sq})$ is merely to change the angle of the gate from $\varphi_\Delta (t_r)$ to $\varphi_\Delta^\text{eff} =\varphi_\Delta (t_r)-\Omega_R t_r$. Furthermore, the effective winding operator takes the form $U_\text{R,eff}(t_{sq}) = U_\text{down} U_{R}(t_{sq})  U_\text{up} =\exp(-i\Omega_R (t_{sq}+t_r) \JJJ_x) $, leading to 
$U_{t_p,0} = U_\text{R,eff}(t_{sq}) U_\text{R,gate}^{\varphi_\Delta^\text{eff} }(t_{sq})$. By applying the unwinding operator $U_\text{R,eff}^\dagger(t_{sq})$ to the the measured density matrix, one recovers the density matrix resulting from the MS gate evolution $ U_\text{R,gate}^{\varphi_\Delta^\text{eff}}(t_{sq})$. 

\begin{figure}[ht]
    \begin{center}  \includegraphics[width=\linewidth]{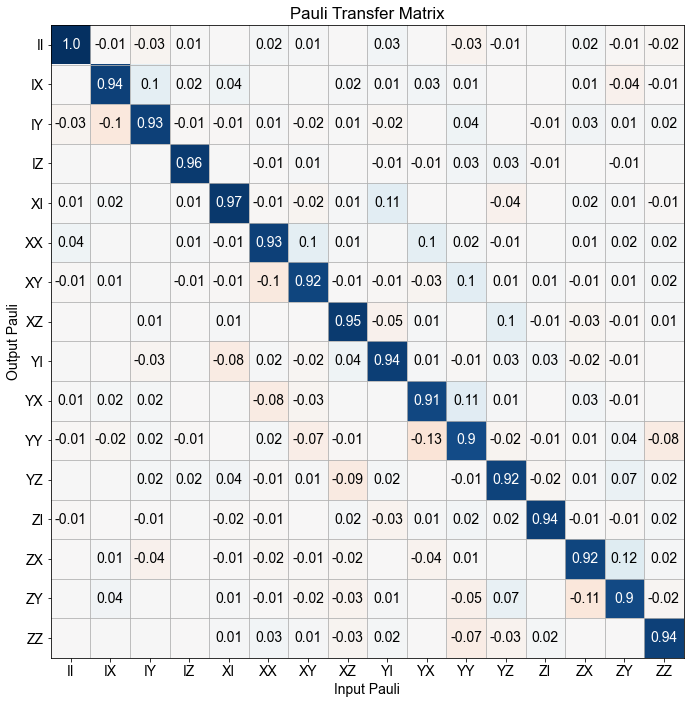}
    \caption{Process tomography of the identity operation}
    \label{fig:ptm_id}
    \end{center}
\end{figure}

\begin{figure*}[ht]
    \begin{center}
        \includegraphics[width=18cm]{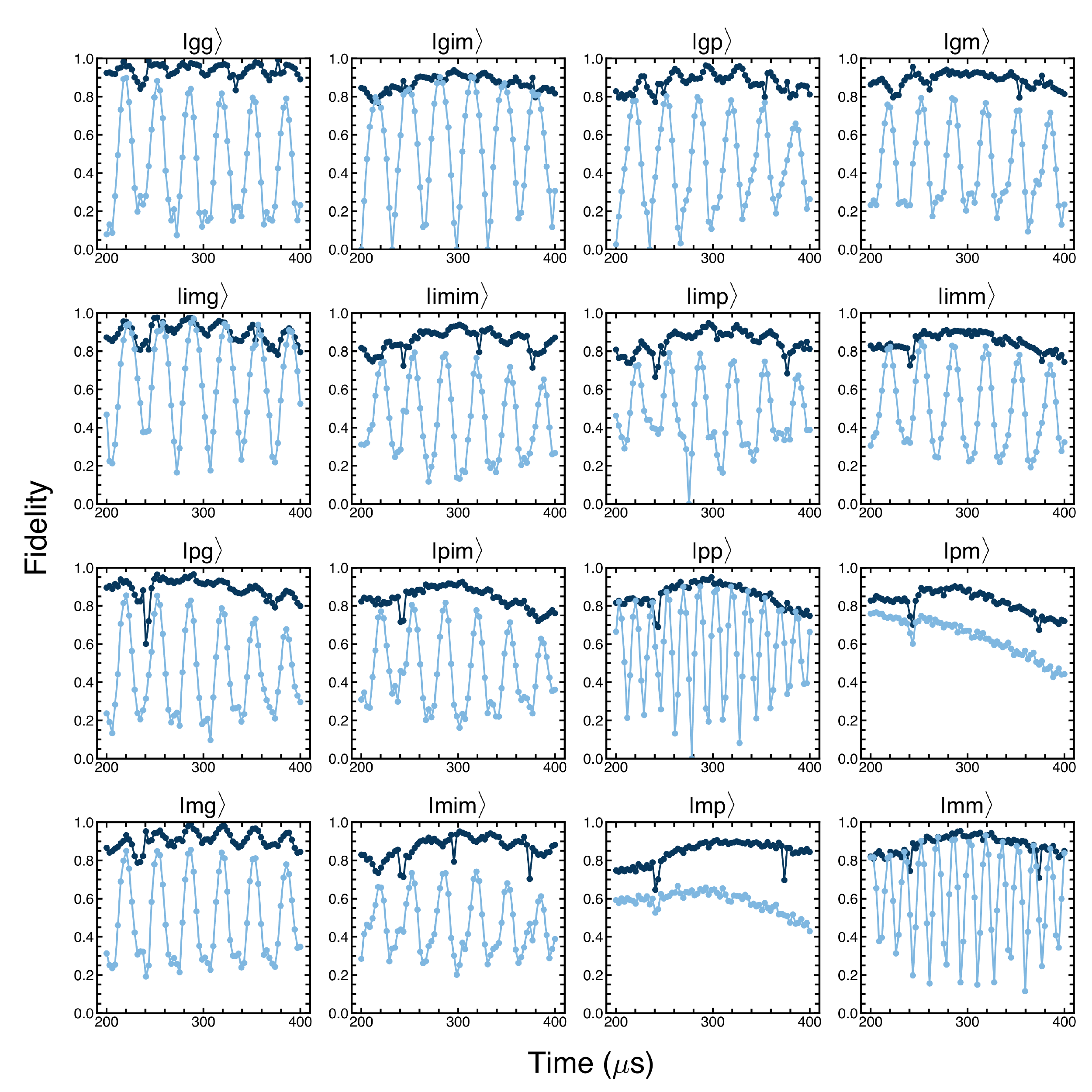}
        \caption{Fidelities to the target states for 16 different initial states labelled in the title of each subplot, used to construct the PTM of figure 3. The light blue points are the fidelities without removing the Rabi drive, exhibiting oscillations at 30 MHz for most initial states, at 60 MHz for $\ket{+,+},\ket{-,-}$ and no oscillations for $\ket{+,-},\ket{-,+}$. The dark blue points are the processed data which have been unwound. The points at t=300 ns give an average fidelity of 92\%.}
        \label{fig:Unwinding_16states}
    \end{center}
\end{figure*}

\subsection{Rabi-driven qubits} \label{sec:sec1}

State of the art fixed frequency qubits still have better lifetimes than flux-tunable frequency ones. Multi-qubit gates however often require precise relations between qubits frequencies, imposing strong requirements on fabrication. We propose and implement here the use of Rabi-driven floquet qubits, with an tunable effective frequency for quantum computation.

Starting with the bare frame Hamiltonian for a driven qubit:

\begin{equation}
    H = -\frac{\omega_q}{2} \sigma_z + \Omega_R \sigma_x
\end{equation}

Going into the frame rotating at the qubit frequency $\omega_z$, or dressed frame, the Hamiltonian of the Floquet qubit is $H_{rot} = \Omega_R \sigma_x$, with two eigenstates: $\ket{+} = (\ket{g} + \ket{e})/\sqrt{2}$ and $\ket{-} = (\ket{g} - \ket{e})/\sqrt{2}$ separated in energy by $2\Omega_R$.

\begin{figure}
    \begin{center}
        \includegraphics[width=9cm]{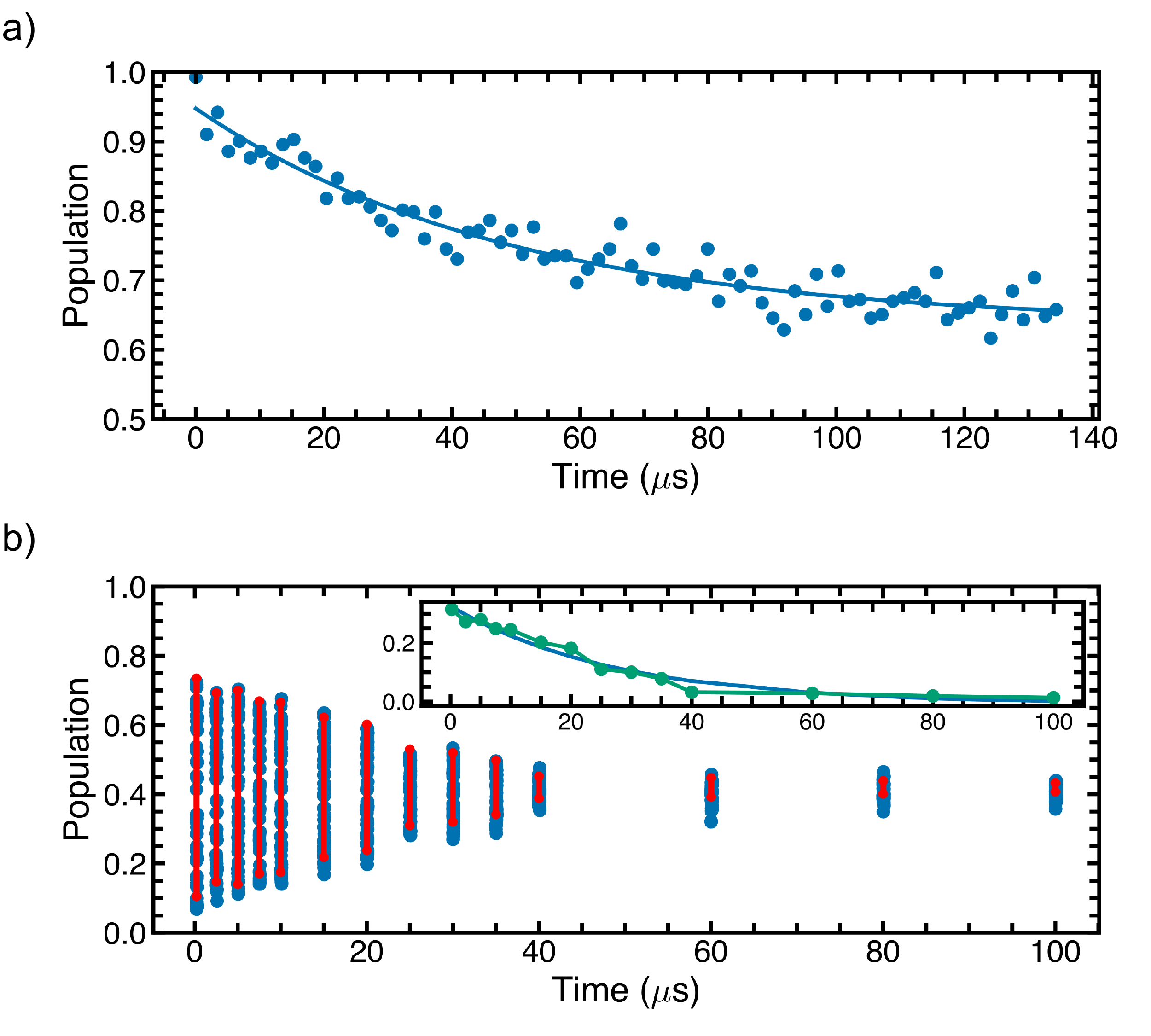}
        \caption{Examples of $T_{1,\rho}$ and $T_{2,\rho}$ measurements. a) A sample $T_{1,\rho}$ measurement obtained from performing a spinlocking pulse sequence. Lifetimes are extracted by fitting an exponential model. b) Example of a $T_{2\rho}$ measurement. We sample a Rabi measurement at different times and extract the max amplitudes (green points in the inset). We  fit the amplitude of Rabi oscillations at different times. Here $T_{2,\rho}= 27 \mu$ s.}
        \label{fig:fig_T2rho}
    \end{center}
\end{figure}

\comment{Rabi-driven qubits have been mainly used in the superconducting qubits community as noise sensors at their rabi frequency \cite{Yan_2013}, allowing to distinguish between coherent and thermal photon noise \cite{Yan_2018, Sung_2021} or to study correlated noise between qubits \cite{von_L_pke_2020}.

The lifetimes in the bare and rotating frame are related by \cite{Sung_2021_noisespec}:
\begin{equation}
    \Gamma_{1,\rho} (\Omega_R) = \frac{1}{2} \Gamma_1 + \Gamma_{\phi} (\Omega_R)
\end{equation}

where $\Gamma_{1,\rho} (\Omega_R)$ denotes the longitudinal spin relaxation rate, and $\Gamma_{\phi} (\Omega_R)$ denotes the relaxation rate due to pure dephasing noise PSD at the locking Rabi frequency $\Omega_R$.

\emph{T1, T2 for 4 qubits}--- Just one iteration for now, we would need more data for error bars. Examples of measurements are shown in \cref{fig:spinlock_q0} and \cref{fig:fig_T2rho}. See table \cref{table_lifetimes} for all the results.}

\begin{table}
    \begin{tabular}{ |c|c|c|c|c|c| } 
    \hline
    Qubit & $T_1$ & $T_{2,Ramsey}$ & $T_{2,echo}$ & $T_{1,\rho}$ & $T_{2,\rho}$\\
    \hline
     0 & 49.3 $\pm$ 14.1  & 13.2 $\pm$ 0.9 & 16.0 $\pm$ 0.9 & 50-60 & 28\\ 
     1 & 57.0 $\pm$ 25.8 & 11.0 $\pm$ 4.7 & 13.9 $\pm$ 2.6 & 50-60 & 16\\
     2 & 48.7 $\pm$ 5.9 & 15.1 $\pm$ 0.5 & 15.7 $\pm$ 0.3 & 60 & 45\\
     3 & 23.8 $\pm$ 3.2 & 11.8 $\pm$ 3.4 & 12.9 $\pm$ 0.5 & 40 & 18\\
    \hline
    \end{tabular}
    \caption{Lifetimes of the 4 qubits without sidebands, all in $\mu$s. The $T_{1,\rho}$ and $T_{2,\rho}$ are shown for a Rabi frequency of 30 MHz.}
    \label{table_lifetimes}
\end{table}

\begin{figure}
    \begin{center}
        \includegraphics[width = \linewidth]{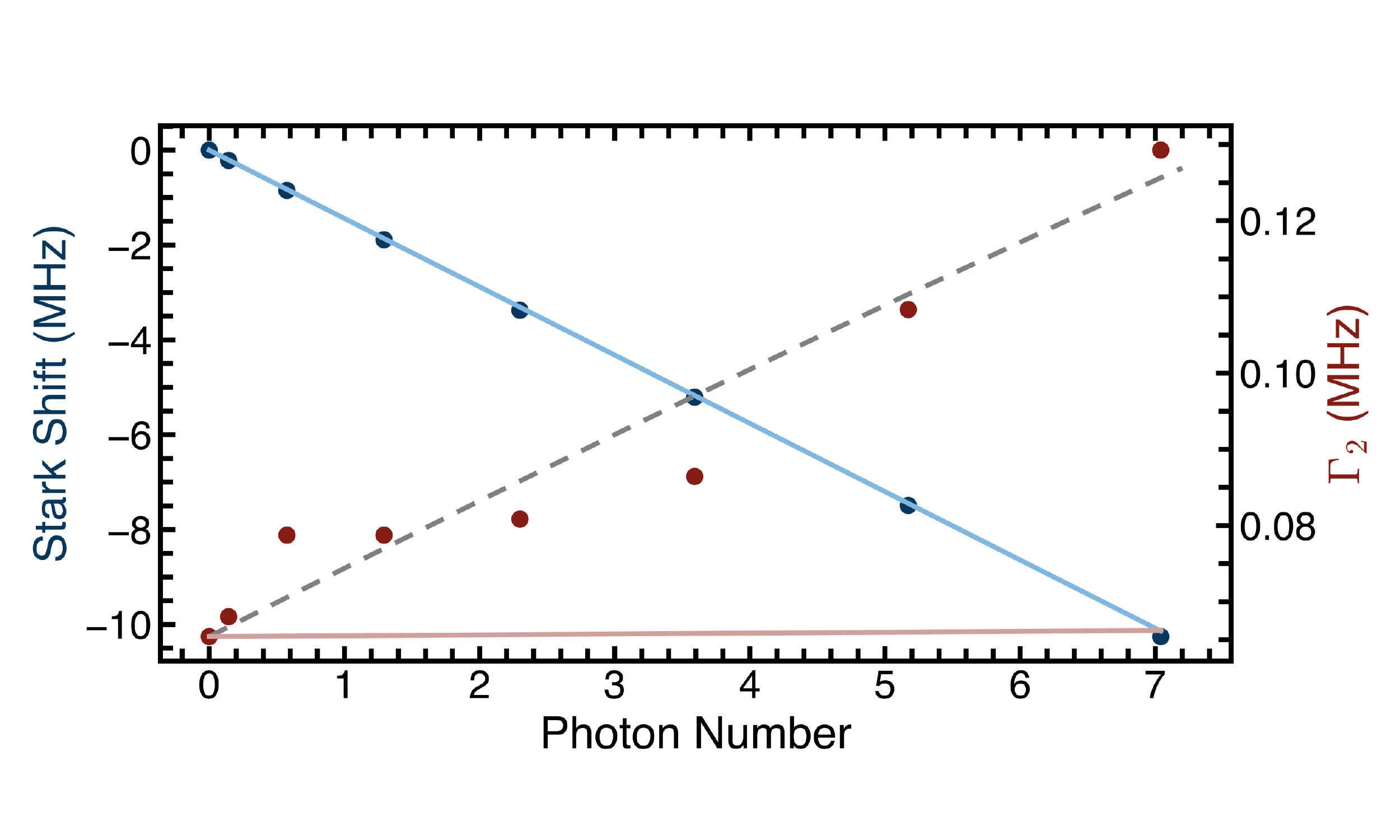}
        \caption{Stark shift and dephasing with 2 sidebands. The blue points correspond to the Stark shift experienced by the qubit. The blue line indicates the linear fit. The red points correspond to the measurement induced dephasing and the red line is the expected dephasing according to \cite{Gambetta_2006}. We see a large difference between the expected from models and actual dephasing. We believe the presence of additional thermals photons could be responsible for this discrepancy and the dashed black line corresponds to the predicted dephasing with the added effect of thermal photons.}
        \label{fig:SShift_Dephasing}
    \end{center}
\end{figure}

\emph{Effect of sidebands on lifetimes}
We perform a study of qubit lifetimes with and without the presence of sidebands and find that the sidebands do not affect the $T_1$ lifetimes but do reduce $T_2$ by 1$\mu$s on average for the sideband strenghts used in the gate. To further characterize this extra loss, we performed a sweep of sidebands power. The resulting Stark shift and dephasing are shown in \cref{fig:SShift_Dephasing}.

\subsection{Single Qubit Pulses quality}

We characterized the pulses via Randomized Benchmarking. The results vary each qubit, ranging from p=0.985 to 0.992 when sidebands are not on and range from 0.98 to 0.985 when sidebands are on. A more complete characterization of the pulse and readout quality would be the PTM for the identity operation, shown in \cref{fig:ptm_id}. The process fidelity for the identity operation is 93.5\%. 

\subsection{Derivation of the MS Hamiltonian from bare frame and driven qubits}

Let us consider two transmon qubits coupled to a resonator through a Jaynes Cumming Hamiltonian, two resonant Rabi drives applied to the transmons and two sideband tones applied to the resonator. The Hamiltonian of the driven system can be written in the following form~\cite{BBQ},
\begin{align}\label{eq:Ham1}
 \mathbf{H}(t)/\hbar&= \tilde{\omega}_{c} \aaa^\dag \aaa+\sum_{k=1}^2 \tilde{\omega}_{k} \bbb_k^\dag \bbb_k \notag\\
& - \sum_{k=1}^2 E_{J}^k \left(\cos\left(\frac{\mathbf{\Phi}_k}{\phi_0}\right)+\frac{1}{2}\frac{\mathbf{\Phi}_k^2}{\phi_0^2}\right)\notag\\
& +\sum_{k=1}^2 \Omega_{R_k}(t)i(\bbb_k^\dag-\bbb_k) \notag\\
& +i(\epsilon_{r}(t)+\epsilon_{b}(t))(\aaa^\dag-\aaa), 
\end{align}
where 
$$
\mathbf{\Phi}_k=\phi^a_{k}(\aaa+\aaa^\dag)+\sum_{k'=1}^2 \phi^b_{k,k'}(\bbb_{k'}+\bbb_{k'}^\dag).
$$
Here we note $\aaa$ (resp. $\aaa^{\dagger}$) and $\bbb_k$ (resp. $\bbb_k^{\dagger}$) the 
annihilation (resp. creation) operator of the resonator and qubit $k$, $\tilde{\omega}_{c}$ and $\tilde{\omega}_{k}$ the dressed frequencies of the resonator and qubit $k$ respectively, $E_{J}^k$ the Josephson energie of qubit $k$, $\phi_0 = \hbar/2e$ the superconducting flux quantum. The drives applied to the resonator (resp. the transmons) also weakly drive the transmons (resp. the resonators) through the hybridization. However, if the frequencies of the system are well separated, these terms can be neglected.

Noting that $\phi^a_{k}\ll\phi^b_{j,j}$, the dressed mode $\aaa$ shares a much smaller part of the non-linearity than the  dressed modes $\bbb_k$. This is why we refer to the $b$ modes as the qubit modes and the $a$ modes as the cavity modes. In the transmon regime, zero-point phase fluctuations of the modes are small, and we can limit the expansion of the cosine to fourth order. Assuming that the frequencies of the system are well separated, and $|\tilde{\omega}_{k}-\tilde{\omega}_{l}| \gg \Omega $, where $\Omega$ is the common Rabi frequency, the Hamiltonian simplifies to 

\begin{align}\label{eq:Ham2}
 \mathbf{H}(t)/\hbar&= {\omega}_{c} \aaa^\dag \aaa+\sum_{k=1}^2 {\omega}_{k} \bbb_k^\dag \bbb_k - K_{k}{\bbb_k^\dag}^2 {\bbb_k}^2 \notag\\
& - \aaa^\dag \aaa \bigg(\sum_{k=1}^2 \chi_k \bbb_k^\dag \bbb_k \bigg) - \sum_{j\neq k}^2 \chi_{j,k} \bbb_j^\dag \bbb_j \bbb_k^\dag \bbb_k \notag\\
& +\sum_{k=1}^2 \Omega_{R_k}(t)i(\bbb_k^\dag-\bbb_k) \notag\\
& +i(\epsilon_{r}(t)+\epsilon_{b}(t))(\aaa^\dag-\aaa), 
\end{align}

where $\omega_{a}$ and $\omega_{k}$ are the renormalized frequencies, $\chi_{k}$, $\chi_{j,k}$ and $K_{k}$ are respectively the cross Kerr coefficient between the resonator and qubit $k$, the cross Kerr coefficient between qubit $j$ and qubit $k$, the self Kerr coefficient of qubit $k$. The Rabi drives and the sideband tones take the forms $\Omega_{R_k}(t) = \Omega_{k}\cos(\omega_k t )$ and $\epsilon_{r/b}(t) = \epsilon_{r}\cos((\omega_c+\Omega_{r/b})t+\varphi_{r/b})$, where $\Omega_{r/b}$ is the detuning of the tone from the resonator frequency.

One can write $\Omega_{r} = -\Omega_{SB} + \delta$, and $\Omega_{b} = \Omega_{SB} + \delta$, where $\delta$ is a small detuning compared to $\Omega_{SB}$.
We eliminate the drive term in the Hamiltonian by applying the displacement on the cavity
\begin{align*}
\aaa =& \ddd + \alpha(t), \notag\\
\alpha(t) =& \frac{\epsilon_{r}e^{-i\varphi_r}}{2i}\bigg(\frac{e^{-i(\omega_{a}-\Omega_{SB} + \delta)t}}{-\Omega_{SB} + \delta}+\frac{e^{i(\omega_{a}-\Omega_{SB} + \delta)t}}{2\omega_{a}-\Omega_{SB} + \delta}\bigg) \notag\\
&+\frac{\epsilon_{b}e^{-i\varphi_b}}{2i}\bigg(\frac{e^{-i(\omega_{a}+\Omega_{SB} + \delta)t}}{\Omega_{SB} + \delta}+\frac{e^{i(\omega_{a}+\Omega_{SB} + \delta)t}}{2\omega_{a}+\Omega_{SB} + \delta}\bigg)
\end{align*}

We choose $\epsilon_b$ and $\epsilon_r$ such that $|\epsilon_b|/(\Omega_{SB} + \delta) = |\epsilon_r|/(-\Omega_{SB} + \delta) = \sqrt{\bar n}$. As $|\epsilon_{r/b}| \ll 2\omega_{a}\pm\Omega_{SB} + \delta$, one can discard the terms having $2\omega_{a}\pm\Omega_{SB} + \delta$ in the denominator. The amplitude of the classical field becomes simply $\alpha(t) = \sqrt{2\bar n} \cos(\Omega_{SB} t +\phi_{\Delta}) e^{-i((\omega_c+\delta) t+\phi_{\Sigma})}$, where $\phi_\Sigma = \frac{\phi_r+\phi_b+\pi}{2}$ and $\phi_\Delta = \frac{\phi_b-\phi_r}{2}$. We drop the phase factor $e^{-i\phi_{\Sigma}}$, as it can be eliminated by the transformation $\aaa \rightarrow \aaa e^{i\phi_{\Sigma}}$. 

Moving to the frame rotating at the frequencies $\omega_{a}+\delta$ and $\omega_{k}$, and discarding the counter rotating terms, the Hamiltonian becomes 

\begin{align}\label{eq:Ham3}
 \mathbf{H}(t)/\hbar&= -\delta \ddd^\dag \ddd-\sum_{k=1}^2 K_{k}{\bbb_k^\dag}^2 {\bbb_k}^2 \notag\\
& - (\ddd^\dag \ddd + \sqrt{2\bar n} \cos(\Omega_{SB} t +\phi_{\Delta})(\ddd^\dag+ \ddd) \notag \\
& +2\bar n \cos^2(\Omega_{SB} t +\phi_{\Delta}) ) \bigg(\sum_{k=1}^2 \chi_k \bbb_k^\dag \bbb_k \bigg) \notag\\
&- \sum_{j\neq k}^2 \chi_{j,k} \bbb_j^\dag \bbb_j \bbb_k^\dag \bbb_k \notag\\
& +\sum_{k=1}^2 \Omega_k \cos(\omega_k t)  i(\bbb_k^\dag e^{i\omega_{k} t }-\bbb_k e^{-i\omega_{k} t }). 
\end{align}

From this more complete form of the Hamiltonian, we note two possible limitations to the performance of the gate. 
First, the cross-Kerr or ZZ interaction, which becomes a resonant XX interaction when the Rabi drives are on. Secondly, another limitation is due to the finite ratio of the Rabi frequency over the anharmonicity of the transmons, $\Omega_k/K_k$.

Assuming that the anharmonicity of the transmons remains larger than the Rabi frequency, i.e $K_k/\Omega_k \gg 1$, we can make a two-level approximations and project the above Hamiltonian on the ground and excited states of the transmons. In addition, we neglect the cross-Kerr between the qubits mediated by the resonator, and perform a rotating-wave approximation on the Rabi drive terms, leading to 

\begin{align}\label{eq:Ham4}
 \mathbf{H}(t)/\hbar&= -\delta \ddd^\dag \ddd+\sum_{k=1}^2 \frac{\Omega_k}{2} \sss_{x_k} \notag\\
& - (\ddd^\dag \ddd + \sqrt{2\bar n} \cos(\Omega_{SB} t +\phi_{\Delta})(\ddd^\dag+ \ddd) \notag \\
& +2\bar n \cos^2(\Omega_{SB} t +\phi_{\Delta}) )\bigg(\sum_{k=1}^2 \tilde \chi_k \sss_{z_k} \bigg), 
\end{align}

where $\tilde \chi_k = \chi_k/2$. In the following and in the main text (under Circuit QED Implementation), we take the definition $\chi_k := \tilde \chi_k$. Setting $\Omega_k = \Omega_{SB}$ and $\chi_k = \chi$ for all $k$, and going into the frame rotating at the Rabi drive frequency, the Hamiltonian becomes that of eq.(\ref{eq:simplifiedHamiltonian}), 

\begin{equation}
    \hhh_R = -\delta \ddd^\dagger \ddd -2\sqrt{2\bar{n}}\chi \JJJ_{zy}^{\varphi_\Delta} (\ddd+\ddd^\dagger)  +\hhh_\text{err}(t),
\end{equation}
where $\JJJ_{\varphi_\Delta} = \cos(\varphi_\Delta)\JJJ_{z}-\sin(\varphi_\Delta) \JJJ_{y}$, $\JJJ_{l} = \sum_k \sss_{l_k}/2,~l=x,y,z$ are the generalized spin operators, and $\hhh_\text{err} = \textbf{A}_1 e^{i\Omega_{SB} t }+\textbf{A}_1 e^{2i\Omega_{SB} t }+\textbf{h.c}$ represents spurious oscillating terms \cite{Eddins2018}. $\hhh_\text{err}(t)$ can be neglected as long as one satisfies $\lVert \textbf{B}_{1,2} \rVert \ll \Omega_{R}$. The dominant term in $\hhh_\text{err}(t)$ comes from the term in third line of eq.~(\ref{eq:Ham4}), and scales as $\chi \bar n e^{i\Omega t } $. This leads to a renormalization of the Rabi frequency that is taken into account in the simulations.  This term is responsible for the fidelity saturation in Fig.\ref{fig:theory_prediction}b. When the Rabi frequency is set to $\Omega_R = 2\Omega_{SB}$, this term becomes resonant and leads to large oscillation of $\langle \sigma_X \rangle$ seen at $\Omega_R = 60~\text{MHz}$ in Fig.~\ref{fig:pulses}c.

\subsection{Master equation}

All simulations are obtained using the following master equation:

\begin{align} \label{eq:masterEq}
    \frac{d\boldsymbol{\rho}}{dt} =& -\frac{i}{\hbar}[\hhh(t), \boldsymbol{\rho}] \notag \\
    &+\sum_k \mathcal{D}\big[\sqrt{{1/2T_{1,\rho,k}}}\sss_{z_k}\big](\boldsymbol{\rho}) \notag \\
    &+\sum_k \mathcal{D}\big[\sqrt{{1/2T_{2,\rho,k}}}\sss_{z_x}\big](\boldsymbol{\rho}) \notag \\
    &+ \mathcal{D}\big[\sqrt{\kappa}\aaa\big](\boldsymbol{\rho}),
\end{align}
where $\mathcal{D}\big[\bf{M}](\boldsymbol{\rho}) = \bf{M} \boldsymbol{\rho} \bf{M}^\dagger - (\bf{M}^\dagger \bf{M}  \boldsymbol{\rho} +   \boldsymbol{\rho} \bf{M}^\dagger \bf{M})/2$.
Here, $T_{1,\rho,k}$ and $T_{2,\rho,k}$ are the spinlocking times of qubit $k$, and we use the Hamiltonian of \cref{eq:Ham4}.

% \begin{figure}[ht]
%     \begin{center}
%         \includegraphics[width=\linewidth]{Figures/Population_Simulation_example.png}
%         \caption{We show the simulated populations as a function of time after initializing the qubits in $\ket{--}$ for couplings and Rabi drives similar to what is present on our processor. The evolution shows that the $\ket{++}$ and the $\ket{--}$ populations meet at around 380 ns. Simultaneously the $\ket{+-}$ and $\ket{-+}$ populations are at their lowest at 240 ns. This should be the time of maximum entanglement.}
%         \label{fig:population_sim}
%     \end{center}
% \end{figure}

\begin{figure}
    \begin{center}
        \includegraphics[width=\linewidth]{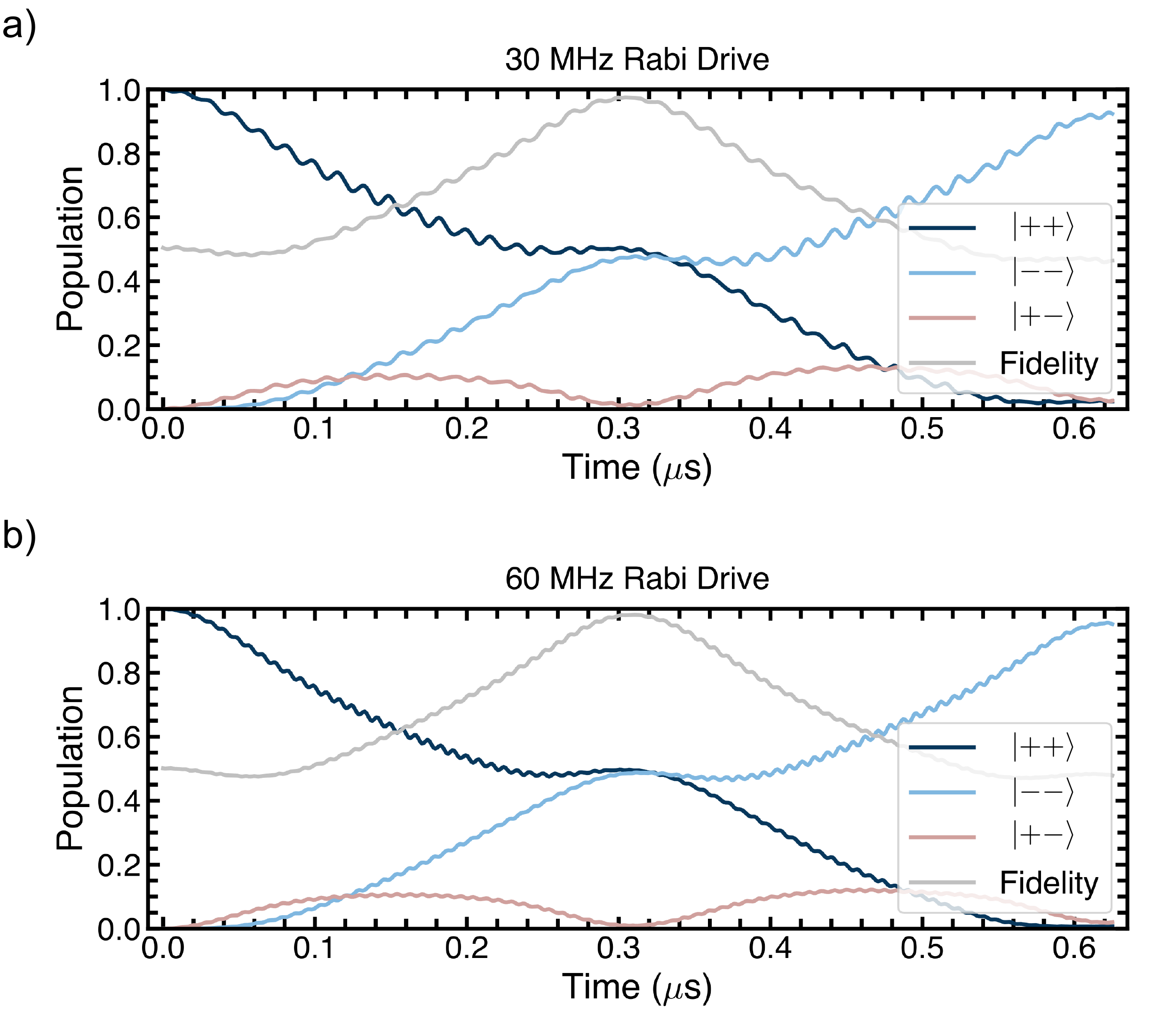}
        \caption{We show the simulated populations as a function of time after initializing the qubits in $\ket{--}$ for couplings and Rabi drives similar to what is present on our processor. These simulations are obtained from simulating \cref{eq:masterEq}. The evolution shows that the $\ket{++}$ and the $\ket{--}$ populations meet at around 300 ns. Simultaneously the $\ket{+-}$ and $\ket{-+}$ populations are at their lowest at 300 ns. This should be the time of maximum entanglement. In addition we show a comparison of population evolution between a gate performed using 30 MHz Rabi drives (top) and 60 MHz Rabi drives (bottom).}
        \label{fig:sim_30vs60}
    \end{center}
\end{figure}

\end{document}